\documentclass[aps,prl,reprint,superscriptaddress, longbibliography]{revtex4-2}

\usepackage{blindtext}
\usepackage{graphicx}
\usepackage{amsfonts}
\usepackage{amsmath}
\usepackage{bbold}
\usepackage{physics}
\usepackage[caption=false]{subfig}
\usepackage{float} 
\usepackage{xcolor}
\usepackage[export]{adjustbox}

\definecolor{dodgerblue}{HTML}{1E90FF}
\definecolor{lightdodgerblue}{HTML}{4dbff7}
\usepackage[colorlinks=true,citecolor=dodgerblue,linkcolor=dodgerblue,urlcolor=dodgerblue,pdftitle={Non-hydro correlators}]{hyperref}

\renewcommand{\l}{\left(}
\renewcommand{\r}{\right)}
\newcommand{\lb}{\left[}
\newcommand{\rb}{\right]}
\newcommand{\lcb}{\left\{ }
\newcommand{\rcb}{\right\} }
\newcommand{\lv}{\left|}
\newcommand{\rv}{\right|}

\newcommand{\mc}[1]{\mathcal{\mathrel{#1}}}
\newcommand{\up}{\uparrow}
\newcommand{\down}{\downarrow}

\renewcommand{\log}[1]{\, {\rm Log} \left[ \mathrel{#1} \right] }

\newcommand{\cbra}[1]{\left(#1\right|} 
\newcommand{\cket}[1]{\left|#1\right)} 

\begin{document}

\title{Subexponential decay of local correlations from diffusion-limited dephasing}

\author{Ewan McCulloch}
\thanks{E.M. and J.A.J. contributed equally to this work.}
\affiliation{Department of Electrical and Computer Engineering, Princeton University, Princeton, NJ 08544, USA}

\author{J. Alexander Jacoby}
\thanks{E.M. and J.A.J. contributed equally to this work.}
\affiliation{Department of Physics, Princeton University, Princeton, New Jersey 08544, USA}

\author{Curt von Keyserlingk}
\affiliation{Department of Physics, King's College London, Strand WC2R 2LS, UK}

\author{Sarang Gopalakrishnan}
\affiliation{Department of Electrical and Computer Engineering,
Princeton University, Princeton, NJ 08544, USA}

\begin{abstract}
Chaotic quantum systems at finite energy density are expected to act as their own heat baths, rapidly dephasing local quantum superpositions. We argue that in fact this dephasing is subexponential for chaotic dynamics with conservation laws in one spatial dimension: all local correlation functions decay as stretched exponentials or slower. The stretched exponential bound is saturated for operators that are orthogonal to all hydrodynamic modes.
This anomalous decay is a quantum coherent effect, which lies beyond standard fluctuating hydrodynamics; it vanishes in the presence of extrinsic dephasing. Our arguments are general, subject principally to the assumption that there exist zero-entropy charge sectors (such as the particle vacuum) with no nontrivial dynamics: slow relaxation is due to the persistence of regions resembling these inert vacua, which we term  ``voids’’. In systems with energy conservation, this assumption is automatically satisfied because of the third law of thermodynamics.
\end{abstract}

\maketitle

\emph{Introduction}.---Chaotic quantum systems at finite energy density are expected to ``thermalize'': i.e., any initial state rapidly evolves into one in which small subsystems have thermal reduced density matrices, possibly with spatially varying temperatures~\cite{mori2018thermalization}. Dynamics beyond this timescale appear to be governed by classical fluctuating hydrodynamics~\cite{forster2018hydrodynamic, spohn2012large, SpohnNLFH, PhysRevX.8.031057, PhysRevX.8.031058, wienand2023emergence}, and the processes by which this description emerges from unitary quantum dynamics has been a topic of intense study~\cite{PhysRevA.89.053608, PhysRevX.8.031057, PhysRevX.8.031058, wienand2023emergence, joshi2022observing, rosenberg2023dynamics, wei2022quantum, le2023observation, gross2017quantum, arute2019quantum, scholl2021microwave, PhysRevA.95.053621, hofferberth2008probing, kitagawa2011dynamics, chiu2019string, rosenberg2023dynamics, 2016Sci...353.1257B, 2020PhRvL.125a0403K, 2022PhRvL.129l3201Y,2023arXiv231213268I, heavyionlecturenotes,heavyionreview, Muller2008,Lucas2016a,Lucas2016b,Crossno2016,Narozhny2017,Lucas2018,Bal2021,PhysRevLett.131.210402,SinghNavierStokes}. The structure of this emergent hydrodynamic description depends only on the symmetries of the system: fluctuations of conserved charges and Goldstone modes relax slowly. More generally the temporal correlations of local operators depend on how they overlap (in a sense we will make precise) with these slow modes, and their products and derivatives, etc. Generally, these correlations decay algebraically with hydrodynamic long-time tails, with an exponent that is fixed by the symmetries of the dynamics and of the operator being considered~\cite{forster2018hydrodynamic, 2023arXiv231010564M, matthies2024thermalization, delacretaz2020heavy}. The details of the underlying  dynamics enter this description only through the values of transport coefficients like the diffusivity; for the purposes of understanding the large-scale behavior of hydrodynamical quantities it is immaterial whether this underlying dynamics is classical or quantum.

Some operators, however, such as the charge-raising operator, have strictly zero overlap with the hydrodynamic modes, and therefore do not have conventional long-time tails. A natural expectation~\cite{Mori_24,2022PhRvB.106v4310N} is that these operators relax as they would in systems with no conservation laws, i.e., exponentially, with a timescale set by the microscopic physics. Intuitively, the charge-raising operator creates coherences between different charge sectors, and one expects such coherences to dephase exponentially rapidly. In the present work we show that this expectation is incorrect: \emph{all} local correlation functions decay subexponentially in systems with a conserved charge, provided there are charge sectors with vanishing entropy. This assumption is always satisfied for Hamiltonian systems, by the third law of thermodynamics. It is also satisfied in charge-conserving systems if there is a unique charge vacuum state.  
Given this assumption, equilibrium states contain rare low-entropy ``voids'' (for example, regions that are locally close to the ground state). A local quantum superposition inserted into a void does not dephase until the void fills in. Thus the rate at which a void fills in, through hydrodynamic processes like diffusion, limits the late-time decay of local correlation functions. We provide evidence that this diffusion-limited dephasing mechanism sets the late-time relaxation of non-hydrodynamic correlation functions in one dimension. 
We find two generic types of behavior, depending on whether the diffusion constant remains finite inside a void: 
if it does (as in random charge-conserving circuits), local correlations decay as $\exp(-\sqrt{t})$; if it instead diverges (as in translation invariant Hamiltonian or Floquet systems), the void fills in faster and local correlations decay as $\exp(-t^{2/3})$. For random charge-conserving circuits, our result is an explicit bound, while for translation-invariant systems, it relies on the (unproven, but generally accepted) assumption that conserved quantities are transported according to the laws of hydrodynamics.

This slow dephasing mechanism is an essentially quantum coherent effect---since dynamics inside voids is coherent---and is unstable to extrinsic dephasing even when that dephasing preserves the conservation laws. By contrast, hydrodynamic transport coefficients are believed to evolve smoothly in the presence of noise that preserves the relevant symmetries~\cite{von_Keyserlingk_2022}; absent any symmetries, the exponential decay rates of correlation functions are also insensitive to weak noise~\cite{Mori_24, Jacoby_25, Zhang_2025}. Since noisy quantum dynamics can be efficiently simulated on classical computers~\cite{aharonov2023polynomial, schuster2024polynomial, gonzalez2024pauli}, a variety of numerical algorithms extract correlation functions by simulating noisy dynamics and extrapolating to the weak noise limit~\cite{Rakovszky_22, von_Keyserlingk_2022, PhysRevB.97.035127} (see also \cite{teretenkov_2025}). Our results suggest, perhaps counterintuitively, that even for local correlation functions these strategies have their limits: stretched exponential relaxation relies in an essential way on maintaining quantum coherence over large scales~\footnote{We note that classical glassy systems~\cite{PhysRevLett.53.958,PhysRevLett.53.1244,PhysRevLett.89.035704,2018PhyA..504..130G} and systems with a long prethermal many-body localized regime~\cite{2023PhRvL.131j6301L,2024PhRvB.109i4207H} also exhibit stretched-exponential decay of correlations despite the absence of any conservation law, but for very different reasons than the diffusion-limited dephasing mechanism we have outlined in this paper.}. As such, it can potentially serve as an experimentally accessible witness of quantum coherence in present-day quantum devices. 

\emph{Background: hydrodynamic overlaps}.---To define more precisely what we mean by non-hydrodynamic operators, we briefly review the concept of hydrodynamic overlaps. The operators acting on a Hilbert space $\mathcal{H}$ form a vector space $\mathcal{O(H)}$, which is equipped with a family of temperature-dependent Bogoliubov inner products. For simplicity we will specialize to the infinite-temperature limit, where the inner product simplifies to $\langle A, B \rangle \equiv \mathrm{Tr}(A^\dagger B)$. In the projection-operator framework for hydrodynamics, one decomposes the operator Hilbert space as $\mathcal{O(H)} = \mathcal{O}_{\mathrm{slow}} \oplus \mathcal{O}_{\mathrm{fast}}$, where $\mathcal{O}_{\mathrm{fast}}$ is the orthogonal complement of $\mathcal{O}_{\mathrm{slow}}$; $\mathcal{O}_{\mathrm{slow}}$ includes not just long-wavelength charge fluctuations but also products of these. To compute the hydrodynamic contribution to the dynamical correlation function of a local operator $A$, one performs a ``hydrodynamic projection'' of $A$ onto the slow subspace. Schematically, to define a hydrodynamic projection of $A$, one fixes a hydrodynamic timescale $\tau$ beyond which the dynamics is said to be slow, then projects the time-average  $\tau^{-1}\int_0^\tau dt A(t)$ into the space $\mathcal{O}_{\mathrm{slow}}$. The details of this procedure are not important for this paper, and arguably remain to be clarified; we refer the reader to Refs.~\cite{doyon2022diffusion, 2019PhRvL.122i1602C, PhysRevB.73.035113, von_Keyserlingk_2022} for background.  The key point for our purposes is that even if $A(0)$ is not in the slow subspace, it can develop overlap with slow operators under time evolution. This is how the current, for example, picks up a projection onto the slow space~\cite{PhysRevB.73.035113}.

In the rest of this work, we will focus on systems of qubits with a single scalar charge, $Q = \sum_x \sigma^z_x$. The operator basis on site $x$ is spanned by the identity operator, $\sigma^z_x$, and $\sigma^\pm_x$, where the latter are spin raising and lowering operators. 
Many-body operators are called \emph{neutral} if they contain equal numbers of $\sigma^+$ and $\sigma^-$, and they are \emph{charged} otherwise. 
Neutral operators connect states in the same charge sector, while charged operators connect states in different charge sectors.
Charge-conserving dynamics also conserves the charge of an operator, so any operator only mixes with others of the same charge. 
The space $\mathcal{O}_{\mathrm{slow}}$ contains only superpositions of products and derivatives of local densities, which are manifestly neutral operators. Under the operator inner product, any charged operator is  therefore manifestly orthogonal to every operator in $\mathcal{O}_{\mathrm{slow}}$, remaining so under time evolution, and is therefore non-hydrodynamic.

\begin{figure}[tb]
    \centering
    \includegraphics[width = \linewidth]{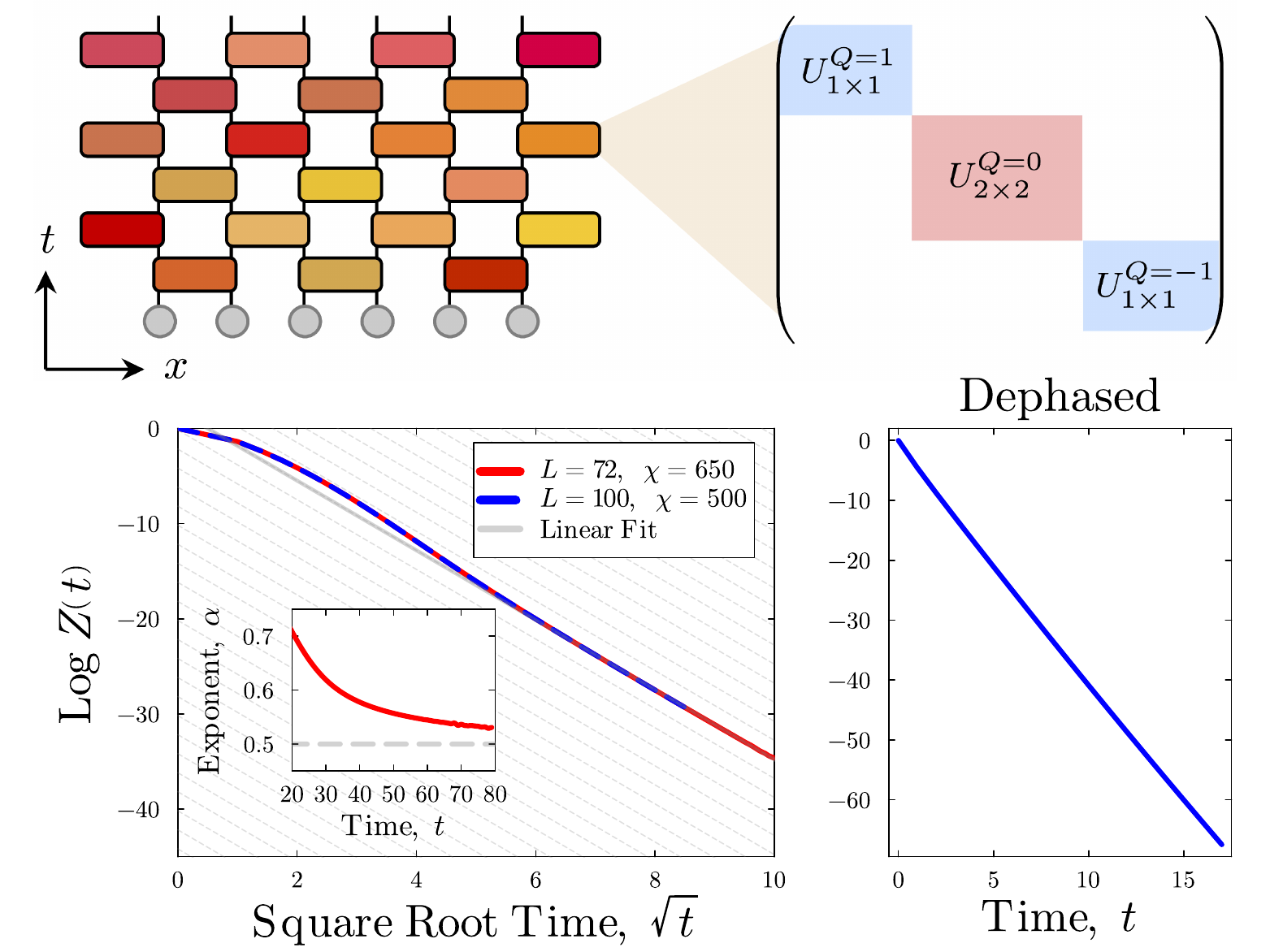}
    \caption{(\emph{Top}) Time-evolution is generated by U$(1)$ conserving two qubit gates arranged in a ``brick wall'' pattern. (\emph{Bottom Left}) Log of position and Haar averaged norm-squared two-point function, $Z\l t \r \equiv \sum_{x}\mathbb{E}_U\lv \langle \sigma^{+}_{0} \l  t \r\sigma^{-}_{x}\rangle \rv^{2}$,   plotted as a function of $\sqrt{t}$. (\emph{Bottom Left, Inset}) Stretching exponent as a function of time (linear scale), ascertained by logarithmic derivative. (\emph{Bottom Right}) $Z(t)$ in the presence of dephasing noise as a function of time (linear scale), showing the breakdown of the stretched exponential behaviour in the presence of extrinsic dephasing. See Ref.~\cite{suppmat} for details of the noisy circuit simulation.}
    \label{fig:u1_blocks}
\end{figure}

\textit{U$(1)$-symmetric random unitary circuits} -- 
We first consider non-hydrodynamic correlation functions in random unitary circuits with a brickwall geometry and a single spin-$1/2$ (qubit) degree of freedom on each site, consisting of block-diagonal gates that conserve total magnetization $Q$ [Fig.~\ref{fig:u1_blocks}]. We investigate the two-point function of the raising/lowering operators $\sigma_x^\pm$, 
\begin{equation}\label{eqn:SM-correlator}
    C_U(x, t) \equiv \langle \sigma^+_x(t) \sigma^-_0\rangle_U \equiv \text{Tr}(U(t)^\dagger\sigma_x^- U(t) \sigma_{0}^+)/\mathcal{Z},    
\end{equation}
where $\mathcal{Z}\equiv\text{Tr}(\mathbb{1})=2^L$. While computing this autocorrelator for a single circuit realization is challenging, the moments of its distribution over circuits $U$---denoted $\mathbb{E}_{U}( |C_U(x,t)|^{2m} )$---can be evaluated using random-circuit techniques~\cite{PhysRevX.8.031058, PhysRevX.8.031057, PhysRevLett.131.210402}. Since the sign of this correlation function will generally oscillate, we estimate its magnitude by computing its second moment, $Z(x,t) \equiv \mathbb{E}_U |C_U(x,t)|^{2}$, which can be expressed in terms of a transfer matrix acting on two replicas of the system~\cite{suppmat}. (Note that averaging over the ensemble of $U$ restores statistical translation invariance in space and time.) Expressed in terms of a two-replica statistical mechanics transfer matrix, we have
\begin{equation}\label{2-rep-corr}
    Z(x,t) = \frac{1}{\mathcal{Z}^{2}} \cbra{\sigma^+_x,\sigma^-_x} T^{t} \cket{\sigma^+_{0},\sigma^-_{0}},
\end{equation}
where $\cket{A,B}$ are vectors in the space of operators acting on two replicas of the Hilbert space, and where we are using the inner product $( A | B ) \equiv \text{Tr}(A^\dagger B)$.  
The transfer matrix $T$ is given by the product of an even and odd layer $T \equiv T_{e} T_{o}$, where $T_{e/o} \equiv \prod_{x \in \text{even/odd}} T_{x,x+1}$,
\begin{equation}\label{transmat}
    T_{x,y} = \mathbb{E}_{U_{x,y}}\left[U_{x,y} \otimes U^*_{x,y} \otimes U_{x,y} \otimes U^*_{x,y} \right].
\end{equation}
Here $\mathbb{E}_{U_{x,y}}$ denotes integration over the Haar measure of a two-site U$(1)$-symmetric random unitary gate. These replica statistical mechanics models are well-established, having been used in the study of scrambling and entanglement growth~\cite{PhysRevX.8.031058, PhysRevLett.122.250602,PhysRevX.8.031057}, full counting statistics~\cite{PhysRevLett.131.210402}, and measurement-induced phase transitions~\cite{PhysRevX.12.041002}. We refer to Ref.~\cite{suppmat} for more details. 

For simplicity we will focus on lower-bounding the local autocorrelation function, $Z(0,t)$. Using the hermiticity of $T$ and inserting a complete orthogonal basis $\{ |v) \}$ for operators in the two-replica space, we obtain
\begin{eqnarray}
    Z(0,t) &\equiv& ( \sigma^+_0,\sigma^-_0 | T^t | \sigma^+_0,\sigma^-_0 ) \nonumber \\
    &=& \sum\nolimits_v | ( v | T^{t/2} | \sigma^+_0,\sigma^-_0 )|^2. \label{variational}
\end{eqnarray}
Since this expression is a sum of positive terms, it is lower-bounded by any of its summands. By choosing an appropriate normalised vector $v$, we can establish the bound.  We choose a state $|v)$ containing $| \sigma^+_0,\sigma^-_0 )$ in the middle of a ``void''---i.e., a fully $z$-polarised region---of size $\ell$, around $x = 0$.

Our choice of state is motivated by the following physical picture which underlies the concrete calculation in  \cite{suppmat}. Inside a void the charged operator $\sigma_0^+$ undergoes single-particle diffusion (up to phase factors) under the U$(1)$ random circuit dynamics. In symmetric random unitary circuits, a domain wall between typical and void regions melts diffusively, over a length-scale ${\cal O}(\sqrt{t})$ by time $t$, and since a single diffusing particle explores a region of space of the same size, a void initially of size $\ell=O(t^{1/2 + \epsilon})$ is sufficient to separate a diffusing $\sigma^+$ operator at its center from the diffusive influx of particles/magnons from the void's edges. In this setup,  the $\sigma^+$ is measured again at the origin after time $t$, and has undergone simple single particle diffusion, which has a slowly (algebraically) decaying amplitude. The more important effect is that voids of length $\ell$ are exponentially rare in the infinite temperature state, appearing with probability $\exp(-{\cal O}(\ell))$. Setting $\ell$ as above, and ignoring the irrelevant subleading algebraic contribution, gives the lower bound
\begin{equation}\label{eqn:bound}
    Z(0,t) \geq 2^{-{\cal O}(t^{1/2 + \epsilon})}, \quad \forall \epsilon > 0.
\end{equation}

Using the time-evolving block decimation (TEBD) algorithm, we simulate the transfer matrix evolution of the correlation function in Eq.~\ref{2-rep-corr} at zero momentum (for improved finite-size behavior) which is expected to be equal to the autocorrelator up to a subleading power law prefactor, and find excellent agreement with a stretched exponential decay $\sum_x Z(x,t) \sim \exp(-{\cal O}(\sqrt{t}))$ (an exactly analogous stretched exponential bound can be derived for this quantity~\cite{suppmat}). This is shown in Fig.~\ref{fig:u1_blocks}. Technically, the argument above only establishes that \emph{some} circuits feature sub-exponential decay of correlations; however, the physical mechanism based on voids clearly applies to any circuit, so we argue that the conclusion should hold for typical circuits. We note that a similar argument involving voids was previously used to show the sub-linear growth of R\'enyi entropies in random charge-conserving circuits \cite{PhysRevLett.122.250602, 2019arXiv190200977H}; our results show that voids also have more direct physical implications.

To verify that the observed sub-exponential decay is indeed a coherent quantum effect, we simulate the random circuit dynamics with dephasing noise, which dephases the operator $\sigma^+$ regardless of whether or not it is embedded in a completely polarized background. We find a clear exponential decay as shown in Fig.~\ref{fig:u1_blocks}. (For details on the noisy circuit simulation, we refer the reader to Ref.~\cite{suppmat}).

\begin{figure*}
    \centering
    \includegraphics[width=\textwidth]{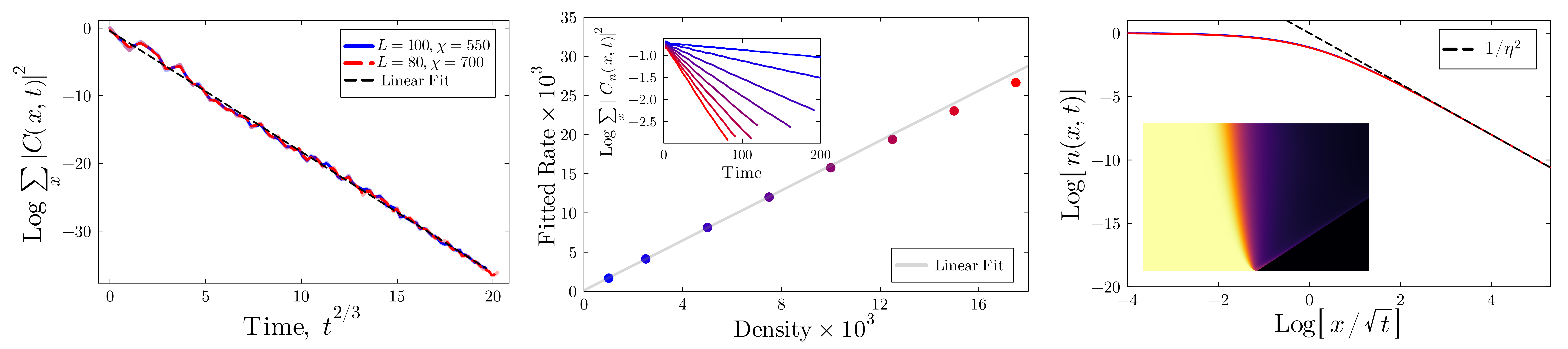}
    \caption{(\emph{Left}) Stretched exponential decay of the non-hydrodynamic correlation function $C(x,t)\equiv\langle \sigma^+_x(t)\sigma^-_0\rangle$ at infinite temperature in a translationally invariant Floquet model with conserved magnetization (model A -- see supplementary material for a definition~\cite{suppmat}). To reduce noise/oscillations, we take the norm-square of the correlation function and sum over $x$.
    We use the time-evolving block decimation (TEBD) algorithm with bond dimensions $\chi=550,700$ and system sizes $L=100,80$ for these simulations (all simulations use the python library TeNPy~\cite{tenpy2024}). See supplementary materials for additional numerical evidence of stretched exponential decay in different models~\cite{suppmat}. 
    (\emph{Center}) The decay rate of $C_n(x,t)$, the local correlation function in a thermal state with magnon density $n$, as a function of $n$ in the low density limit, showing clear exponential decay with a rate proportional to magnon density. As with the infinite temperature (half-filling) case, we take the norm-square of the correlator and sum over $x$ to reduce noise. We use TEBD with bond dimension $\chi=200$ and system sizes $L\approx 400$ and simulate to times $t={\cal O}(1)$ in units of the mean-free-time. (\emph{Center, Inset}) $\sum_x |C_n(x,t)|^2$ as a function of time at various densities (represented by the color scale, see main panel), smoothed with a Gaussian kernel with standard deviation $\Delta t=2.5$ (in units of the Floquet period).
    (\emph{Right}) Scaling collapse of the density $n(x, t)$ as a function of $\eta=x/\sqrt{t}$ on a log-log plot for an initial domain wall configuration with $\mc{O}\l 1 \r$ charge density for $x<0$ and a void for $x>0$. The dynamics are from a two-species interacting lattice gas with $D\sim1/n$ \cite{suppmat}, used to model void melting. The key feature of the collapse is strong agreement in the regime of $\eta \sim t^{1/6} \to \infty  $, which is addressed further in \cite{suppmat}. Numerical parameters are $L=4\cdot 10^{4}$ and time samples at multiples of $\delta t  = 1000$. All data has converged in system size. (\emph{Right, Inset}) Relaxation of a domain wall with high density in yellow and low density in purple/black. A ballistic front, followed by the scaling regime can be observed in the bottom half of the figure; the ballistic front is responsible for a parametrically small amount of charge transport. A diffusive melting regime can be observed slowly encroaching upon the scaling collapse regime, which appears black due to the low overall density. }
    \label{fig:2}
\end{figure*}

\emph{Translationally Invariant Floquet Systems}.---
In the rest of this paper, we turn from random circuits to Floquet systems---i.e., those with a time-periodic Hamiltonian $H(t) = H(t+1)$---that are also translation invariant in space.  These additional symmetries give rise to stable ballistically propagating charged excitations in the zero-density limit, and this crucially modifies the physics of voids and how they fill in. We argue that, nevertheless, autocorrelation functions decay as stretched exponentials, but with a different stretching exponent,  $C(t) \geq \exp(-\mc{O}(t^{2/3}))$, where in the present section we focus on the simple local autocorrelator $C(t) \equiv \langle \sigma^{+}_0(t)  \sigma^{-}_0 \rangle$. 

We begin with two observations: (i) Even if transport deep inside a void is ballistic, particle interactions in the higher density regions at the edge of the void limit the rate at which particles can enter the void. We show this using standard hydrodynamic reasoning below. (ii) Dephasing is slow even in imperfect voids \cite{Chen_2020_1,Chen_2020_2}: the correlator $C(t)$ in a region of density $n$ decays no faster than $\exp(-\mc{O}(n) t)$. Intuitively, this is because the operator $\sigma^{+}_{0}$ inserts a magnon into a low-density region, and this magnon dephases through randomly timed collisions with background particles, which occur at a rate set by the background density. The numerical evidence for this linear density dependence of dephasing is clear, see Fig.~\ref{fig:2}. 

We use these observations to upper bound the decay rate as follows. Starting with a state that contains a void of size $\ell$ centered at the origin, we apply the operator $\sigma^+_0$, creating a magnon. Within the void, the magnon encounters a density profile $ n(x,t)$ which is, before the void fills in, much smaller than the typical density outside the void. If the void were perfectly empty, the magnon would propagate as a free particle, and its return probability, and the corresponding contribution to $C(t)$, would decay as a power law in time~\footnote{This is true for generic band structures in which each band has zero average group velocity. It fails for topological Floquet band structures with certain sublattice symmetries (and our conclusions do not apply to such models).}.  When the inserted magnon encounters a background magnon, the two get entangled and (using (ii)) each such collision damps the survival probability of the magnon. We will show that it is voids of size $\ell\sim t^{2/3}$ that contribute most to the correlator at time $t$. This scaling arises from striking a balance between two competing effects: as $\ell$ gets larger, the likelihood of an initial void of size $\ell$ decreases; on the other hand, larger voids more effectively protect the magnon from dephasing.

To proceed, we estimate the background charge distribution, starting from an initial void of size $\ell$. In general, two-body collisions on the lattice can relax momentum, so the mean free path scales as $1/n$. Coarse-grained over length-scales $\gg 1/n$, the density dynamics can therefore be described by a diffusion equation with a density-dependent diffusion constant $D(n) \approx D_0/n$. We use these properties to write the phenomenological transport equation (see also \cite{Chen_2020_2,Zanoci_2021})
\begin{equation}
    \partial_t n = D_0\partial_{x} \left(n^{-1} \partial_{x} n \right) .
    \label{eqn:ld_transport}
\end{equation}
Setting $D_{0} = 1 $ for simplicity, we find scaling solutions to this equation of the form $n=F(\eta)$ where $\eta = x \ t^{-1/2}$ and where $F F'' + \frac{1}{2}F' \l \eta F^{2} - 2F'\r = 0$. This equation is solved by $F\l \eta \r\sim \eta^{-2}$ in the low density limit. The density profile can be compared to numerical direct simulations of the hydrodynamics per Fig.~\ref{fig:2}.

Equipped with this solution for $ n(x,t)$, we first roughly estimate the contribution of the rare region to our correlator before producing a more detailed model which reproduces our naive prediction. In the center of the void, and before the void has filled in, we predict a density of $ \propto t/\ell^2$. Combining the probability of the initial void with the density dependent bound on the decay of the correlator in the void (ii) gives $|C(t)| \gtrsim e^{-a_{1} \ell} e^{- ( a_{2} t/\ell^2) \times t}$, where $a_{1/2}$ are real constants. We now choose the void size $\ell$ so as to optimize the lower bound. We find that $\ell \propto t^{2/3}$, which results in a $\exp(-t^{2/3})$ (or slower) decay of the correlator, as claimed above. This scaling is consistent with our assumption that the void is large enough that it does not fill in prior to time $t$, this is diffusion limited and thus should take a time $\sim \ell^2 =O( t^{4/3}) \gg t$.  Our analytical solution $1/\eta^{2}$ is a likely an overestimate of the scaling of the central density since the dynamics become more diffusive over time, with the implication that $C(t)$ might decay even slower than we predict. Using TEBD simulations, we calculate the correlation function $C(x,t)=\langle \sigma^+_x(t) \sigma^-_0\rangle$ for several chaotic translationally invariant Floquet model with conserved magnetization (see~\cite{suppmat} for more details) and find a stretched exponential decay in excellent agreement with our conjectured bound on the stretching exponent $\alpha=2/3$. This is shown in Fig.~\ref{fig:2}.

One might worry that the above calculation neglects the spatial and temporal fluctuations in the density; the following more detailed model attempts to account for that physics. Late-time contributions to the correlation function come from quantum trajectories where the magnon avoided scattering with any background particle. The magnon propagator, conditioned on avoiding collisions, solves the single-particle non-Hermitian Hamiltonian
\begin{equation}\label{nhh}
H_{\mathrm{n.h.}}
(t) = \sum_{x,y} J_{xy}(t) \ket{x}\hspace{-0.15mm} \bra{y} + i \gamma(t) \sum_x  \hat n(x,t) \ket{x}\hspace{-0.15mm} \bra{x}, 
\end{equation}
where $J_{xy}(t)$ is a time-periodic hopping Hamiltonian, and $\gamma(t)$ is an (also time-periodic) interaction strength. In the above equation, $\hat n(x,t)$ is a  background density, subject to stochastic diffusive dynamics. (We use hats to distinguish the fluctuating variable $\hat n$ from its coarse-grained average $n$.) We find numerically that a quantum particle subject to these dynamics has a survival probability scaling as $\exp(-t^{2/3})$ and, neglecting density fluctuations, derive a lower bound with the same scaling. We refer the reader to Ref.~\cite{suppmat} for more details on the numerical simulation and the analytic bound (including a derivation of Eq.~\eqref{nhh} using a projected path integral formalism).

We note that an exactly parallel argument can be used to bound the scaling of half-system R\'enyi entropies after a quench: for R\'enyi index $\alpha > 1$, these entropies scale at most as $t^{2/3}$ in translation-invariant systems. Previous work included a rigorous proof of $t^{1/2}$ scaling in systems where the diffusion constant remains finite at low density \cite{2019arXiv190200977H}; our (weaker) bound applies even when the diffusion constant diverges.

\emph{Hamiltonian systems}.---We now briefly comment on the case of Hamiltonian systems. When these systems have conserved charges beside energy, we can define charged operators exactly as above, and the same arguments give stretched-exponential decay with a stretching exponent $\alpha = 2/3$. When the only conserved charge is the energy itself, two potential subtleties arise. First, there is no simple construction of non-hydrodynamic local operators. This does not invalidate our bound, which holds for \emph{any} local operator, but might render it vacuous at very late times unless such non-hydrodynamic operators exist. Second, the nature of zero-temperature transport is not universal, as it depends on the nature of the ground state. For the maximally generic case of a nondegenerate ground state with a gapped, quadratic quasiparticle dispersion above it, the arguments above seem to apply, giving the same bound $\alpha = 2/3$. However, in the presence of additional lattice or internal symmetries, more exotic ground states can arise; to adapt our arguments to these settings, we would need to incorporate the details of low-temperature energy transport. 

\emph{Discussion}.---The main result of this work is that, under generic conserving dynamics, the dephasing of local correlations is subexponential in time, even when these correlations are orthogonal to all hydrodynamic variables. For random charge-conserving circuits, we have substantiated our argument with an explicit statistical mechanics calculation. For generic translation-invariant systems, we have provided an argument based on standard beliefs about hydrodynamics and operator growth. In both cases, we have presented unambiguous numerical support for our main claims. The diffusion-limited dephasing mechanism poses a challenge for certain approximate numerical methods~\cite{Rakovszky_22, von_Keyserlingk_2022, PhysRevB.97.035127} to study quantum dynamics: can these methods be adapted to preserve coherent dynamics in {the components of the wavefunction which contain} voids?

It would be interesting to extend our results to systems with discrete symmetries, and to higher dimensions. A naive application of our arguments to higher dimensions suggests that the contributions from voids decay as  $\exp(-t^{2d/(d+2)}),\exp(-t^{d/2})$ in the RUC/Floquet cases respectively. A more thorough investigation of this question is required, however, particularly in the `critical' case of $d=2$, where these contributions first become exponentially decaying. Another direction is to investigate the possibility of \emph{subdiffusion}-limited dephasing in constrained systems. A final outstanding task is to reconcile these subexponential decays with the concept of Ruelle-Pollicott resonances~\cite{Mori_24, prosen2002ruelle, PhysRevE.110.054204}. In systems without conservation laws, these resonances (associated with the exponential decay of correlations) exist as well-defined eigenvalues of the quantum channel in the weak dissipation limit. With a conservation law, as we have seen, correlations do not decay exponentially. Whether this implies the absence of Ruelle-Pollicott resonances, or a more subtle connection between these and local correlation functions, is a question for future work.

\emph{Acknowledgements}.--- The authors thank David Huse, Carolyn Zhang, and Juan P. Garrahan for helpful discussions. E.M. and S.G. acknowledge support from the Co-design Center for Quantum Advantage (C2QA) under contract number DE-SC0012704. J.A.J. was partially supported by the National Science Foundation Graduate Research Fellowship Program under Grant No. DGE-2039656. C.K. is supported by a UKRI FLF MR/Z000297/1. The simulations presented in this article were performed on computational resources managed and supported by Princeton Research Computing, a consortium of groups including the Princeton Institute for Computational Science and Engineering (PICSciE) and the Office of Information Technology's High Performance Computing Center and Visualization Laboratory at Princeton University. Any opinions, findings, and conclusions or recommendations expressed in this material are those of the authors and do not necessarily reflect the views of the U.S. National Science Foundation.

\bibliography{Refs}

\pagebreak

\widetext

\newpage

\makeatletter
\begin{center}
\textbf{\large Supplemental Materials: Subexponential decay of local correlations from diffusion-limited dephasing}

\vspace{3mm}

Ewan McCulloch,\textsuperscript{1} J.~Alexander~Jacoby,\textsuperscript{2} Curt von Keyserlingk,\textsuperscript{3} Sarang Gopalakrishnan\textsuperscript{1}

\vspace{2mm}

\textsuperscript{1}\textit{\small Department of Electrical and Computer Engineering,\\ Princeton University, Princeton, NJ 08544, USA}

\textsuperscript{2}\textit{\small Department of Physics, Princeton University, Princeton, New Jersey 08544, USA}

\textsuperscript{3}\textit{\small Department of Physics, King's College London, Strand WC2R 2LS, UK}

\makeatother


\makeatother

\end{center}
\setcounter{equation}{0}
\setcounter{figure}{0}
\setcounter{table}{0}
\setcounter{page}{1}
\makeatletter
\renewcommand{\theequation}{S\arabic{equation}}
\renewcommand{\thefigure}{S\arabic{figure}}

\section{U$(1)$-symmetric random unitary circuits}

In this section we introduce the replica statistical mechanics model used to compute circuit averaged moments of observables in U$(1)$-symmetric random unitary circuits~\cite{PhysRevX.8.031058, PhysRevX.8.031057, PhysRevLett.131.210402, PhysRevLett.122.250602}, and then use this model to bound the late time decay of a class of correlation functions that are outside the predictions of hydrodynamics. We consider random unitary circuits with a brickwork geometry as shown in Fig.~\ref{fig:u1_blocks}. We take the gates to be U$(1)$-symmetric, conserving the total $z$-component of spin, enriching the random circuit with a diffusive conserved charge. This conservation law is enforced by the block diagonal structure of the unitary gates given in Fig.~\ref{fig:u1_blocks}. A single time-step involves the application of a layer of unitary gates on the even-odd bonds, followed by an application of a layer of unitary gates acting on odd-even bonds, so that the unitary at time step $t$ is given by $U_t = U_{t,O} U_{t,E}$, with $U_{t,E/O} = \prod_{x\in \text{Even}/\text{Odd}} U_{t;(x,x+1)}$, where $U_{t;(x,x+1)}$ is a two-site U$(1)$-symmetric random unitary matrix. We denote the full unitary evolution operator as $U(t)\equiv U_t \cdots U_1$. Heisenberg evolution of an observable $O$ is then given by $O(t) \equiv U(t)^\dagger O U(t)$, and connected (auto)correlation functions of $O$ in an initial mixed state $\rho$ are given by 
\begin{equation}\label{correlation function}
    \langle O^\dagger O(t) \rangle^c_{\rho} \equiv \Tr(\rho O^\dagger O(t)) - \Tr( \rho O^\dagger)\Tr(\rho O(t) ).
\end{equation}
Rather than focus on any individual random circuit realization, we will consider the circuit average of the moments of these correlation functions. The first moment (the average) of correlations with charged operators $O$ always vanishes, $\mathbb{E}_U O(t) = 0$, where $\mathbb{E}_U \left[ X \right]$ denotes the circuit average of the quantity $X$. To see this, let $O$ have a definite charge $\Delta Q$, i.e., the only non-zero matrix elements $\bra{a} O \ket{b}$ are those where the states $\ket{a}$ and $\ket{b}$ differ by a definite charge $\Delta Q$. This means that when written as a linear combination of the strings of $\mathbb{1},\sigma^z,\sigma^\pm$, the imbalance in the number of $\sigma^+$'s and $\sigma^-$'s is equal to $\Delta Q$. When applying, for example, the even-odd layer of the unitary evolution to such an operator, one must always find at least one even-odd pair of sites on which the operator has an imbalanced number of $\sigma^\pm$. Without loss of generality we focus on the case where the imbalance is positive, so that either a single $\sigma^+$ is present alongside a neutral operator $A=\mathbb{1},\sigma^z$, or two $\sigma^+$'s are present. The local update is then given by
\begin{equation}\label{local-operator-update}
    U_{x,y}^\dagger \sigma^+_x A_y U_{x,y} = \sum_{Q=-1,0} \left(U^{Q+1}_{x,y}\right)^\dagger \sigma^+_x A_y \left(U^{Q}_{x,y}\right), \quad U^\dagger_{x,y} \sigma^+_x \sigma^+_{y} U_{x,y} = \left(U^{Q=1}_{x,y}\right)^\dagger \sigma^+_x \sigma^+_y \left(U^{Q=-1}_{x,y}\right),
\end{equation}
where $U_{x,y}=\sum_{Q}U^Q_{x,y}$ is a two-site U$(1)$-symmetry random gate, with different symmetry block indexed by $Q=0,\pm1$. Because the charged operators $\sigma^+ A_y$ and $\sigma^+_x \sigma^+_y$ select different blocks of the unitary $U_{x,y}$ and its hermitian conjugate $U^\dagger_{x,y}$, the Haar average of Eq.~\eqref{local-operator-update} is zero (since each block has an independent random phase). This gives $\mathbb{E}_U(O(t))=0$ for all charged operators $O$, requiring us to look at the higher moments of correlation functions involving $O$ to the characteristic decay of non-hydrodynamic operators in symmetric random unitary circuits. In the following we will restrict our analysis to second moments, but the analysis naturally generalizes to higher moments.

\subsection{Replica statistical mechanics model}

Since the observable $O$ is charged by assumption, we can discard the disconnected part of the correlation function. Utilizing the trace inner-product $\left(A|B\right) \equiv \Tr(A^\dagger B)$, we can rewrite the correlation function as $\langle O^\dagger O(t)  \rangle_\rho = \left(O \rho|O(t)\right)$. The `states' $\cket{A}$ are vectorized operators, living on an enlarged Hilbert space defined through the isomorphism $\ket{a}\bra{b} \to \cket{\ket{a}\bra{b}} \equiv \ket{a}\otimes \ket{b^*}$. In this representation, the correlation function is a matrix element of the `doubled' unitary operator, $\langle O^\dagger O(t)\rangle_\rho = \cbra{O \rho}U(t)\otimes U(t)^*\cket{O}$, where the tensor product is between two copies of the original Hilbert space -- the forward and backwards Keldysh contours. This can be seen using the isomorphism once again $ U\ket{a}\bra{b}U^\dagger \to U\ket{a}\otimes U^*\ket{b^*} = \left(U\otimes U^*\right) \ket{a} \otimes \ket{b^*}$. We now restrict our analysis to the second moment $|\langle O^\dagger O(t) \rangle_\rho|^2$, the circuit average of which is given by

\begin{equation}
    \mathbb{E}_U|\langle O^\dagger O(t)  \rangle_\rho|^2 \equiv \cbra{ O \rho,\rho O^\dagger} \mathbb{E}_U \left[U(t)\otimes U(t)^*\otimes U(t) \otimes U(t)^* \right]\cket{O,O^\dagger}.
\end{equation}
The circuit average of the replicated unitary can be interpreted as a transfer matrix, $T(t) \equiv \mathbb{E}_U \left[U(t)\otimes U(t)^*\otimes U(t) \otimes U(t)^* \right]$, which inherits the brickwork geometry of the random unitary circuit, $T(t)=T^t$, $T=T_O T_E$, where $T_{E/O} = \prod_{x \in\text{Even}/\text{Odd}} T_{x,x+1}$. The two-site transfer matrix is computed in Refs.~\cite{PhysRevX.8.031057} and~\cite{von_Keyserlingk_2022}, and is given by
\begin{align}
    T_{x,y} = &\ \cket{I_{1,1}}\cbra{I_{1,1}}+\cket{I_{-1,-1}}\cbra{I_{-1,-1}} + \cket{I_{1,-1}}\cbra{I_{1,-1}}+\cket{I_{1,-1}}\cbra{I_{1,-1}}\nonumber\\
    & + \frac{1}{2}\left(\cket{I^+_{1,0}}\cbra{I^+_{1,0}}+\cket{I^+_{-1,0}}\cbra{I^+_{-1,0}} + \cket{I^+_{0,1}}\cbra{I^+_{0,1}}+\cket{I^+_{0,-1}}\cbra{I^+_{0,-1}} \right)\nonumber\\
    & + \frac{1}{2}\left(\cket{I^-_{1,0}}\cbra{I^-_{1,0}}+\cket{I^-_{-1,0}}\cbra{I^-_{-1,0}} + \cket{I^-_{0,1}}\cbra{I^-_{0,1}}+\cket{I^-_{0,-1}}\cbra{I^-_{0,-1}} \right)\nonumber\\
    & + \frac{1}{3}\left(\cket{I^+_{0,0}}\cbra{I^+_{0,0}}+\cket{I^-_{0,0}}\cbra{I^-_{0,0}}\right) - \frac{1}{6}\left(\cket{I^+_{0,0}}\cbra{I^-_{0,0}}+\cket{I^-_{0,0}}\cbra{I^+_{0,0}}\right),
\end{align}
where the states $\cket{I_{q_1,q_2}}$ for $q_1,q_2 \in \{-1,1\}$, are given by $\cket{I_{2s,2s'}} \equiv \cket{P_s,P_{s'}}_x\cket{P_s,P_{s'}}_y$ where $P_{\up/\down}$ is the projector onto the $\up/\down$ spin, and where the states $\cket{I^\pm_{q,0}}$ for $q=\pm1$ are given by
\begin{align}
    &\cket{I^+_{1,0}} \equiv \cket{P_\up,P_\up}_x\cket{P_\up,P_\down}_y + \cket{P_\up,P_\down}_x\cket{P_\up,P_\up}_y\quad &&\cket{I^+_{-1,0}} \equiv \cket{P_\down,P_\up}_x\cket{P_\down,P_\down}_y + \cket{P_\down,P_\down}_x\cket{P_\down,P_\up}_y\nonumber\\
    &\cket{I^-_{1,0}} \equiv \cket{P_\up,P_\up}_x\cket{\sigma^-,\sigma^+}_y + \cket{\sigma^-,\sigma^+}_x\cket{P_\up,P_\up}_y \quad &&\cket{I^-_{-1,0}} \equiv \cket{P_\down,P_\down}_x\cket{\sigma^+,\sigma^-}_y + \cket{\sigma^+,\sigma^-}_x\cket{P_\down,P_\down}_y.
\end{align}
The states $\cket{I^\pm_{0,q}}$ are found by swapping the sites $x\leftrightarrow y$. Finally, the states $\cket{I^\pm_{0,0}}$ are defined by
\begin{align}
    &\cket{I^+_{0,0}} \equiv \cket{P_\up,P_\up}_x \cket{P_\down,P_\down}_y + \cket{P_\down,P_\down}_x \cket{P_\up,P_\up}_y + \cket{P_\up,P_\down}_x \cket{P_\down,P_\up}_y + \cket{P_\down,P_\up}_x \cket{P_\up,P_\down}_y \nonumber\\
    &\cket{I^-_{0,0}} \equiv \cket{P_\up,P_\up}_x \cket{P_\down,P_\down}_y + \cket{P_\down,P_\down}_x \cket{P_\up,P_\up}_y + \cket{\sigma^+,\sigma^-}_x \cket{\sigma^-,\sigma^+}_y + \cket{\sigma^-,\sigma^+}_x \cket{\sigma^+,\sigma^-}_y.
\end{align}
Note that the statistical mechanics model state space is only six dimensional on each site, since the transfer matrix annihilates any state with  $\cket{\sigma^\pm,\sigma^\pm}$, $\cket{\sigma^\pm ,P_s}$ , or $\cket{P_s, \sigma^\pm }$ on any site. This is the same statistical mechanics model used in~\cite{von_Keyserlingk_2022} to clarify the role of operator backflow processes in hydrodynamics. We will now use it to study correlation functions beyond hydrodynamics.

\subsection{Bounding charged operator decay with rare void regions}

So far we have kept the charged observables $O$ completely generic. For simplicity, we now focus on the simplest example, that of a single raising operator $\sigma^+_x$. However, this restriction can be easily relaxed, and any manner of non-hydrodynamic observable can be dealt with similarly. Our aim is to understand the decay of $\sigma^{+}_x$ through the correlation function $\langle \sigma^+_x(t)\sigma^-_{x'} \rangle_\rho$ using the replica statistical mechanics model introduced above. Here we study the circuit averaged second moment of the correlation function, $\mathbb{E}_U |\langle \sigma^+_x(t) \sigma^-_{x'}\rangle_\rho|^2$, at infinite temperature, $\rho\propto \mathbb{1}$, and define
\begin{equation}\label{2-rep corr}
    Z(x,t)\equiv \mathbb{E}_U |\langle \sigma^+_x(t) \sigma^-_0\rangle_{T=\infty}|^2 = \frac{1}{\mathcal{Z}^2}\cbra{\sigma^+_y,\sigma^-_y} T^t \cket{\sigma^+_x,\sigma^-_x},
\end{equation}
where $\mathcal{Z} \equiv \Tr(\mathbb{1})$ includes the normalization of the infinite temperature state. By moving into momentum space, the circuit averaged second moment can be expressed as a diagonal matrix element of the transfer matrix, $Z(k,t) = \bra{k}T^t \ket{k} $, where $\cket{k}\equiv \frac{1}{\sqrt{L}} \sum_x e^{ikx} \cket{\sigma^+_x,\sigma^-_x}$. Using the hermiticity of the transfer matrix, $T = T^{\dagger}$, and inserting a resolution of identity, we write $C(k,t)$ as
\begin{equation}
    Z(k,t) = \frac{1}{\mathcal{Z}^2}\sum_{\ket{v}}\cbra{k} T^{ \frac{t}{2} \dagger}\cket{v} \cbra{v} T^{ \frac{t}{2}} \cket{k},
\end{equation}
where $\{|v\rangle \}$ is an orthonormal basis. Since every term in the sum is manifestly non-negative, each bound the correlation function from below. Denoting  $B_v(k,t) \equiv \cbra{v}T^{ \frac{t}{2}} \cket{k}/\mathcal{Z}$, we have $Z(k,t) \geq \sum'_v |B_v(k,t)|^2$, for a sum $\sum'_v$ over any subset of the basis vectors $\cket{v}$. Our bound is achieved by choosing a particular subset of $L$ orthonormal vectors $\{\cket{v_y}\}$ indexed by a position $y$. Note that this is a tiny fraction of the total state space which is exponential in system size. For the vectors $\cket{v_y}$, we make the choice $\cket{v_y}\propto\cket{\sigma^+_y \rho_{R_y}, \rho_{R_y} \sigma_y^-}$ where $\rho_{R_y}$ is a density matrix prepared with the fully polarized state $\ket{\downarrow\cdots \downarrow}\bra{\downarrow\cdots \downarrow}$ in a region $R_y$ (the ``void") with radius $r$ centered on site $y$, and at infinite temperature outside of $R_y$ as shown below
\begin{equation}
    \rho_{R_y} \propto \underbrace{\mathbb{1}\cdots \mathbb{1}}_{y-r-1}\ \underbrace{P_\down \cdots P_\down}_{r} \ P_\down\ \underbrace{P_\down \cdots P_\down}_{r}\ \underbrace{\mathbb{1} \cdots \mathbb{1}}_{L-r-y},
\end{equation}
and similarly for $\sigma^+_y\rho_{R_y}$ and $\rho_{R_y}\sigma^-_y$ (leaving the site indices implied),
\begin{align}
    &\sigma^+_y\rho_{R_y} \propto \mathbb{1}\cdots \mathbb{1} \ P_\down \cdots P_\down \ \sigma^+\ P_\down \cdots P_\down\ \mathbb{1} \cdots \mathbb{1}\nonumber\\
    &\rho_{R_y}\sigma^-_y\propto \mathbb{1}\cdots \mathbb{1} \ P_\down \cdots P_\down \ \sigma^-\ P_\down \cdots P_\down\ \mathbb{1} \cdots \mathbb{1}.
\end{align}
For $\cket{v_y}$ to be normalized, we finally have $\cket{v_y}=2^{L-l_R}\cket{\sigma^+_y \rho_{R_y}, \rho_{R_y} \sigma_y^-}$ where $l_R = 2r+1$ is the size of the rare region. 

Since U$(1)$-symmetric random unitary circuits have a density independent diffusion constant $D(\mu)=D$ for all fillings/chemical potentials $\mu$, the fully polarized void survives for a time $\tau\sim l_R^2/D$, being melted diffusively from its edges. Until this time, the interior of the void is dynamically frozen, except for the single excitation created by the raising operator at the center of the void which undergoes single particle diffusion under the dynamics generated by $T$. In the remainder of this section, we will assume that we are taking the thermodynamic limit $L\to\infty$ before the long time limit. Using translational invariance, we have
\begin{equation}
    B_{y}(t) = \frac{2^{-l_R}}{\sqrt{L}}e^{iky} \sum_{x'} e^{-ikx'} \cbra{\sigma^+_{x'} \rho_{R_{x'}}, \rho_{R_{x'}} \sigma_{x'}^-} T^{ \frac{t}{2}} \cket{\sigma_0^+, \sigma_0^-}.
\end{equation}
In a totally polarized background, the two-replica operators $\sigma_{x'}^+\otimes \sigma_{x'}^-$ remain bound and undergo a random walk as a single particle (since the transfer matrix annihilates the state if they separate onto different sites), exploring a region of space of size ${\cal O}(\sqrt{Dt})$. After a time $t$, the domain-wall between the rare-region and the infinite temperature state smears over a length $\ell = {\cal O}(\sqrt{Dt})$. By letting the rare-region size scale as $l_R\sim t^{\alpha}$, $\alpha > 1/2$, we parametrically separate the smeared domain-walls and the random walk of $\sigma_{x'}^+\otimes \sigma_{x'}^-$, so that as $t\to\infty$, the random walker undergoes single particle diffusion (without any many-body collisions) with probability $1$. At time $t/2$ the contraction with with $\cket{\sigma_0^+, \sigma_0^-}$ forces the random walk to have an end point at $x=0$. Furthermore, with our choice $\alpha>1/2$, the relaxation of the domain walls occurs without interactions from $\sigma_{x'}^+\otimes \sigma_{x'}^-$, allowing us to write $B_{y}(t)$ as 
\begin{equation}
    B_{y}(t) = \frac{2^{-l_R}}{\sqrt{L}} e^{iky} \sum_{x'} e^{-ikx' }P(x',t/2) \cbra{\rho_{R_{x'}}, \rho_{R_{x'}}} T^{t/2} \cket{\mathbb{1},\mathbb{1}},
\end{equation}
where $P(x,t)$ is the probability that a random walk starting at the origin finishes at a position $x$ after a time $t$. To arrive at the above equation, we have assumed that the origin lies within a diffusive cone of $x'$, $|x'|\sim \sqrt{t}$ (and so lying within the void). We are able to make this assumption as contributions from larger $x'$, i.e., $x'\sim t^\beta$, $\beta>1/2$, decay no slower than the super-polynomial decay $P(x\sim t^\beta,t)\sim e^{-t^{2\beta -1}}$, which vanishes as $t\to\infty$. Using the asymptotic Gaussian form for the distribution $P(x,t)$, evaluating the sum over $x'$ gives $\sum_{x'} e^{-ikx'} P(x',t/2) = \exp(-Dk^2 t)$. Furthermore, using the fact that the maximally mixed state is a stationary state of $T$ ($T\cket{\mathbb{1}, \mathbb{1}}=\cket{\mathbb{1}, \mathbb{1}}$) and also the fact that $\text{Tr}(\rho_{R_{x'}})=1$ to write $\left(\rho_{R_x'},\rho_{R_x'}|\mathbb{1}, \mathbb{1}\right)=1$, we find $B_{y}(t) = 2^{-{\cal O}(t^\alpha)}/\sqrt{L}\times \exp(-Dk^2t)$. Since the vectors $\cket{v_y}$ are orthogonal, we can achieve an improved bound by summing over $y$. This gives,
\begin{equation}
    Z(k,t)\geq \sum_y |B_y(t)|^2 = 2^{-{\cal O}(t^\alpha)} e^{-Dk^2t}, \quad \alpha > 1/2.
\end{equation}

\subsection{Exponential decay of correlations from $\sigma^z$-dephasing noise}
\begin{figure}
    \centering
    \includegraphics[width=0.43\linewidth]{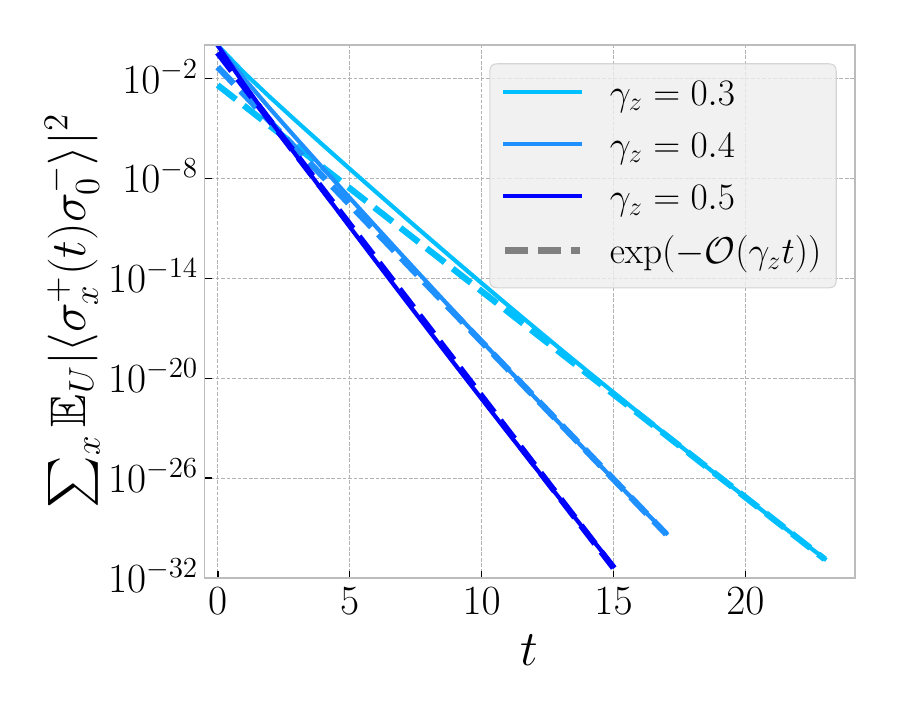}
    \caption{Exponential decay of the non-hydrodynamic correlation function $\langle \sigma^+_x(t)\sigma^-_0\rangle$ (norm-squared and summed over $x$, and averaged over circuits) in random unitary circuits with $\sigma^z$-dephasing noise with noise strength $\gamma_z=0.2,0.3,0.4$. Unlike the isolated case, non-hydrodynamic correlation functions are observed to decay exponentially in time. We use TEBD for these simulations with bond dimension $\chi=500$ and system size $L=100$.}
    \label{fig:RUC-dephasing-noise}
\end{figure}

In this section, we consider U$(1)$-symmetric random unitary circuits with $\sigma^z$-dephasing noise. Just as in the closed system, this open system dynamics has a single diffusive conserved charge (total magnetization). However, unlike in the isolated case, the charged operator $\sigma^+$ cannot be protected from dephasing events by charge voids, since the external noise dephases this operator regardless of the local charge density. The evolution we consider is a layer of random unitary gates (as in the isolated case), followed by a layer of dephasing gates, implementing the Lindblad evolution $\partial_t O = \hat{\mathcal{D}}(O)$ for a timestep $\Delta t=1$ for an operator $O$. The dissipator $\hat{\mathcal{D}}$ is defined by
\begin{equation}
    \hat{\mathcal{D}}(O) \equiv \gamma_z\sum_x \Big[L_x O L_x^\dagger - \frac{1}{2}\{L_x^\dagger L_x, O\} \Big],
\end{equation}
where $L_x=\sigma^z_x$ are Lindblad jump operators for $z$-dephasing and $\gamma_z$ is the dissipation strength. We simulate the circuit-averaged dynamics of the correlation function $\sum_x |\langle \sigma^+_x(t)\sigma^-_0\rangle|^2$ using the TEBD algorithm with bond dimension $\chi=500$ and system size $L=100$ for noise strengths $\gamma_z = 0.3,0.4,0.5$, and find clear exponential decay. This is shown in Fig.~\ref{fig:RUC-dephasing-noise}. We show the noisy circuit simulation with noise strength $\gamma_z=0.4$ in Fig~\ref{fig:u1_blocks} of the main text.

\section{U$(1)$ Transport Model for translationally invariant systems}

Per the main text, using the fact that the mean-free-path of a particle at small density in translationally invariant Floquet systems is inversely proportional to density~\cite{Chen_2020_2}, we consider a diffusive fluid with diffusivity $D\l n \r\sim 1/n$. This yields the equation $\partial_t n=\partial_{x} \left(n^{-1}\partial_{x} n\right)$, which can be solved in the diffusive scaling regime at low density by $n \sim t/x^2$. However, in the main text we consider a beyond-diffusive regime with $x\sim t^{2/3}$ rather than $x\sim t^{1/2}$, so this scaling approach is a priori an uncontrolled approximation. It is therefore important to verify the scaling form holds in this regime by direct simulations of the hydrodynamics. We make use of a discrete-time nonlinear Markov chain to simulate the dynamics. The results are shown in Fig.~\ref{fig:2} of the main text.

Beginning with two vectors of length $L$ (systems size) which represent two local densities, $n_L$ and $n_R$, the time evolution consists of two parts: (i) a translation by one lattice site left/right (ensuring that the model has ballistic transport at zero density) (ii) a density dependent interaction which converts between the two (and which conserves only the total density $n\equiv n_L+n_R$). Defining $T_{L/R}= \sum_{x}\ket{x\mp 1 }\bra{ x }$, and leaving the $x$ dependence implicit, step (i) is given by
\begin{equation}
    n_{L}\l t + 1/2\r = T_{L} \  n_{L}\l t \r \qquad  n_{R}\l t + 1/2\r = T_{R} \ n_{R}\l t \r.
\end{equation}
For the interaction term (step (ii)) we have, at each position,
\begin{eqnarray}
    n_{L}\l t +1\r &=& \frac{\lambda \,  n_{L}\l t + \frac{1}{2} \r + \sigma \,  n_{R} \l t + \frac{1}{2} \r}{1 +  \lambda \,  n_{L} \l t + \frac{1}{2} \r + \sigma \,  n_{R}\l t + \frac{1}{2} \r} \ n_{R}\l t + \frac{1}{2} \r+ \frac{1}{1 +  \lambda \,  n_{R}\l t + \frac{1}{2} \r + \sigma \,  n_{L}\l t + \frac{1}{2} \r}\ n_{L}\l t + \frac{1}{2} \r \nonumber \\
    n_{R}\l t + 1 \r &=& \frac{\lambda \,  n_{R}\l t + \frac{1}{2} \r + \sigma \,  n_{L}\l t + \frac{1}{2} \r }{ 1+\lambda \,  n_{R}\l t + \frac{1}{2} \r + \sigma \,  n_{L}\l t + \frac{1}{2} \r }\ n_{L}\l t + \frac{1}{2} \r + \frac{1}{1 +  \lambda \,  n_{L}\l t + \frac{1}{2} \r + \sigma \,  n_{R}\l t + \frac{1}{2} \r} \ n_{R}\l t + \frac{1}{2} \r.
\end{eqnarray}
This step ensures that the integrated density is conserved, positive semidefinite, and transport is ballistic in the zero density limit. The parameter $\lambda$ can be considered an intraspecies interaction strength and $\sigma$ an interspecies interaction strength. Under these dynamics, particles change species, and therefore velocity, at an ${\cal O}(n)$ rate, giving the desired diffusivity $D(n)\sim 1/n$. For the right panel of Fig.~\ref{fig:2}, we make use of the parameters $\lambda = 2$ and $\sigma = 1$ in the main panel and $\lambda = 1$ and $\sigma = 0.01$ (with $\sigma$ taken very small to emphasize the ballistic front) in the inset.

\begin{figure}[h]
    \centering
    \includegraphics[width=0.7\linewidth]{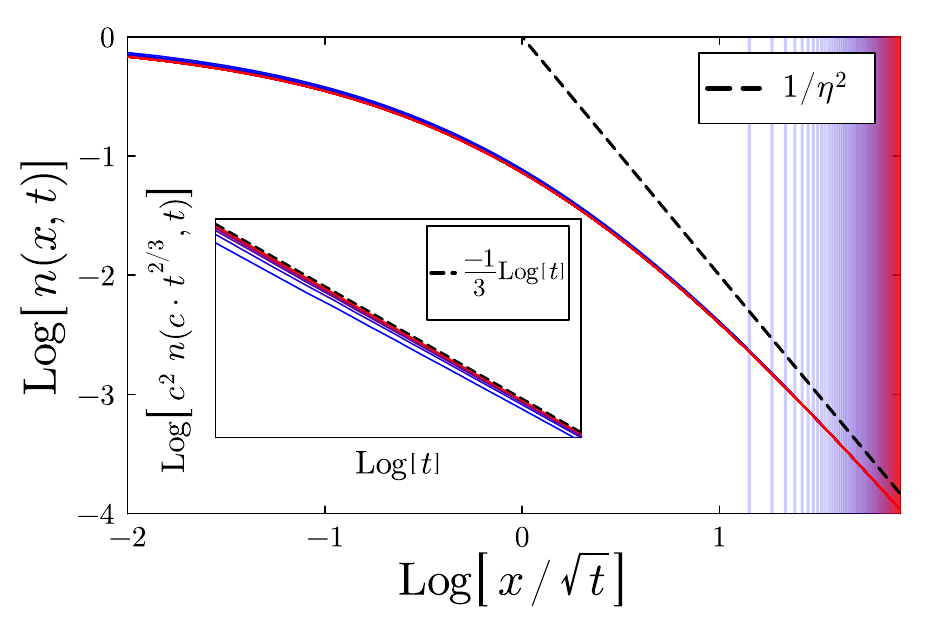}
    \caption{The diffusive scaling collapse of $n(x,t)$ with $t$ ranging from $10^{3}$ to $10^{5}$ in increments of $\delta t = 10^{3}$; different times are indicated by color coding (cold to hot denoting early to late times). For each of these times we denote the point $x = t^{2/3}$ with a vertical line of the same color. In this scaling regime the rescaled coordinate $\eta=x/\sqrt{t}$ diverges as $t^{1/6}$ (moving along the scaling collapse in the $+x$ direction with increasing $t$) and the density becomes arbitrarily close to the scaling solution $n \sim \eta^{-2}$. Per the main text, the model parameters are $\lambda = 2$ and $\sigma = 1$. The system size is $10^5$. (Inset) $n(x,t)$ plotted at fixed $x = c \cdot t^{2/3}$ against $t$ on a log scale showing an asymptotic $t^{-1/3}$ decay with increasing $t$. Vertical offsets are accounted for by scaling $n$ by a factor of $c^{2}$ since $n \sim \frac{t}{x^{2}}\to \frac{t}{c^{2}x^{2}}$. A range of $c$'s are shown, from $2$ to $8$ (again, cold to hot).}
    \label{fig:2/3_collapse}
\end{figure}

Returning to the regime of $x\sim \ell \sim t^{2/3}$, we can straightforwardly see that $\eta \sim t^{1/6}$ and $n\l \ell, t \r \sim t^{-1/3}$ in good agreement with numerical simulations (see inset of Fig.~\ref{fig:2/3_collapse}). The choice of $\alpha$ for $\ell\sim t^{\alpha}$ for $ 1/2 < \alpha \lesssim 1 $ does not appear to make a difference insofar as agreement with the scaling solution. When $x\sim t^{\alpha}$ with $ 1/2 < \alpha \lesssim 1 $, $\eta \sim t^{\alpha - 1/2}$ will diverge. We are therefore most interested in the agreement of the scaling collapse with the prediction $\eta^{-2}$ as $\eta \to \infty$; here we see excellent agreement with numerics. 

\section{Magnon Path Integral and Imaginary Effective Potential}

\subsection{Derivation of Magnon Path Integral}

In this subsection we will re-derive the dissipative single magnon evolution of the main text (Eq.~\eqref{nhh}) from a more microscopic perspective and analytically bound the correlation function under a number of simplifying assumptions. We begin with the standard correlation function in the infinite temperature state
\begin{equation}\label{corr}
    \langle \sigma^{-}_{x_{t}}\l t\r\sigma^{+}_{x_{0}}\l 0 \r \rangle = \sum_{\psi \in \mc{H}_{\rm V}} p\l \psi \r\bra{\psi} \sigma^{-}_{x_{t}}\l t\r\sigma^{+}_{x_{0}}\l 0 \r \ket{\psi} + \sum_{\psi\in \mc{H} \setminus \mc{H}_{\rm V}}   p \l \psi \r \bra{\psi}\sigma^{-}_{x_{t}}\l t\r\sigma^{+}_{x_{0}}\l 0 \r \ket{\psi}
\end{equation} 
where we have separated out states $\ket{\psi}\in \mc{H}_{\rm V}$ which contain a void of size $\ell$ around the raising/lowering operator, from those that do not. The void size $\ell$ will be later be optimized to find the dominant contribution to the correlation function. We now conjecture that the second term is subleading, and ignore it henceforth. 
In general, one could split up the time evolution into a path sum by inserting a complete set of states at each time step. We instead insert an \emph{incomplete} set of states---those consisting of a single magnon above the instantaneous ``background'' state that would obtain if $\ket{\psi}$ had been allowed to evolve without any perturbation. The histories that are left out of this prescription are those in which the magnon interacts nontrivially with the background; such histories do not contribute to local correlation functions. This no-backflow assumption \cite{von_Keyserlingk_2022} neglects autocorrelation of the background density. Following these steps, we arrive at the following ``path integral'' representation of the correlation function:
\begin{equation}
    \bra{\psi} \sigma^{-}_{x_{t}}\l t\r\sigma^{+}_{x_{0}}\l 0 \r \ket{\psi} 
    = \bra{\psi \l t \r} \sigma^{-}_{x_{t}} U^{t}\sigma^{+}_{x_{0}}\ket{\psi\l 0 \r}
    \approx \bra{\psi \l t \r} \sigma^{-}_{x_{t}} \prod_{\tau : \, t-1 \leftarrow 1  } \l U  \sum_{x_{\tau}}   \sigma^{+}_{x_{\tau}}\ket{\psi\l \tau \r} \bra{ \psi\l \tau\r} \sigma^{-}_{x_{\tau}}\r  U \sigma^{+}_{x_{0}}\ket{\psi}.
\end{equation}

We can reorganize this path integral by evaluating the single-timestep, state-dependent propagator from $x$ to $y$:
\begin{eqnarray}
    &&\bra{\psi \l  \tau \r} \sigma^{-}_{y}U\sigma^{+}_{x}\ket{\psi\l \tau - 1\r}  = \bra{\psi \l  \tau \r} \sigma^{-}_{y}U\sigma^{+}_{x}U^{\dagger}\ket{\psi\l \tau \r} \nonumber \\
     & &\ \ \approx e^{- \gamma \bra{\psi\l \tau \r}\hat{n}\l \frac{x+y}{2}\r\ket{\psi\l\tau\r} } G_{0}\l x, y \r \braket{\psi\l\tau \r} = e^{- \gamma \bra{\psi\l \tau \r} \hat{n}\l \frac{x+y}{2}\r\ket{\psi\l\tau\r} } G_{0}\l x, y \r
\end{eqnarray}
where $G_{0}\l x, y \r = \bra{y} e^{-i\omega\l \hat{p}\r}\ket{x}$ is the free magnon propagator and $\gamma$ is a model dependent parameter. 
To approximate this expectation value, we adopt a mean field assumption: set $n(x,t) \equiv \bra{\psi(t)} \hat{n}(x)\ket{\psi(t)}$, where $n\l x, t\r$ is a classical ``background'' density field over which the magnon will evolve.
%
%
In the continuous-time limit, $\hat U_{n\l t \r}$ is the propagator associated with the annealed non-Hermitian Hamiltonian in Eq.~\eqref{nhh} of the main text (after reinstating time-dependent couplings which are periodic in the Floquet period). 

The mean density approximation allows us to analytically lower bound the survival probability of the magnon and obtain an estimate on the size of the correlation function using the annealed single-body magnon propagator $\hat{U}_{n(\tau)} = \sum_{x,y}\ket{y}e^{-\gamma n\l \frac{x+y}{2}, \tau\r} G_{0}\l x, y \r \bra{x}$. In doing so, we approximate the correlation function in states $\psi$ with a void of length $\ell$ around the raising/lowering operators as $\langle \sigma^{-}_{y}\l t\r\sigma^{+}_{x}\l 0 \r \rangle \sim p_{\ell} \bra{y} \prod_{\tau: t \leftarrow 1 } \hat{U}_{n\l \tau \r} \ket{x}$
where $p_{\ell} \sim e^{- \mc{O}\l \ell\r}$ is the probability of finding such a void. Using the above equation we can rewrite the norm-squared correlator (summed over positions) as
\begin{equation}
    \ell^{-1}\sum_{x,y} \lv \langle\sigma^{-}_{y}\l t \r \sigma^{+}_{x}\rangle\rv^{2} \sim \ell^{-1} p_{\ell}\sum_{x,y} \Big| \bra{y}\prod_{\tau: t \leftarrow 1 } \hat{U}_{n\l \tau \r} \ket{x}\Big|^{2} > \ell^{-1}p_{\ell}\sum_{x,y} \Big| \bra{y} \l \hat{U}_{n(t)}\r^{t}\ket{x}\Big|^{2} , 
    \label{eqn:bound_start}
\end{equation}
where in the last inequality we have assumed the density at all times to be the same as the density at late times, overestimating dissipative effects. In the next section we will lower bound the right hand side of Eq.~\eqref{eqn:bound_start}, accounting for the magnon's velocity and kinetic energy; as a consequence we will also control subleading terms in the density profile in a perturbation series in $x/\ell$ where $x$ is the center of the void (the local minimum of the background density).

\subsection{Harmonic Oscillator Solutions and Subleading Terms}

Our aim in this section is to lower-bound the right hand side of Eq.~\eqref{eqn:bound_start} in terms of the eigenvalues of a non-Hermitian operator. We proceed as follows: first, we expand the magnon propagator around minima of the background density and group velocity in position and momentum space respectively. This allows us to write $\hat U_{n(t)}$ (in what follows we will suppress the subscript and denote this as $U$) as the evolution of a harmonic oscillator with an imaginary potential. We can rewrite the RHS of Eq.~\eqref{eqn:bound_start} as the sum of the propagator's singular values squared, which can be lower bounded using a well known inequality which relates functions of the singular values and functions of the absolute values of eigenvalues (since the generator of the propagator is highly non-normal the singular values and eigenvalues are not equal). This demonstrates that the naive scaling of the void correlation function's decay, as $t^{2}/\ell^{2}$, is correct and not overwhelmed by subleading terms resulting from interplay between the kinetic energy of the magnon and the background density (decay profile). Ancillary to this point, the form of the eigenstates shows that the magnon remains in the void and the higher order terms in the expansion of the background density will not contribute on dimensional grounds.

We begin from the magnon propagator (dropping subscripts and explicit time dependence): $\hat{U} =\exp\l -i \lb \omega\l\hat{p}\r - i n\l \hat{x}, t \r \rb t \r $. We can expand $n\l\hat{x},t\r$ around its minimum as $n(\hat{x},t)\approx\frac{1}{8}t\big[ \l \ell/2 + \hat{x} \r^{-2} + \l \ell/2 -  \hat{x} \r^{- 2}\big]  = \frac{t}{\ell^{2}}  + \frac{12 t}{\ell^{4}} \hat{x}^{2} + \, ... $ where $1/8$ is added for convenience and we set $\gamma = 1$. We can then rewrite the propagator as
\begin{equation}
    \hat{U} = \exp\Big(  - i \frac{\hat{p}^{2}}{m^{*}} - i \omega_{0} -t/\ell^{2} - k \hat{x}^{2} \Big) = \exp\l -i \epsilon\l \hat{p} , \hat{x}\r\r
\end{equation}
where $k = 12 t/\ell^{4} \sim t^{1-4\alpha}$ and we project onto a local minimum of the dispersion so that the magnon's group velocity is minimized. We will drop $\omega_{0}$ as it does not survive the norm of Eq.~\eqref{eqn:bound_start}.

Though this operator is non-normal, we can derive the eigenstates and quasispectrum \cite{Davies_99,Davies_99_2}. Working in the coordinate basis,
\begin{equation}
    \epsilon\l \hat{p}, \hat{x}\r = e^{-i\pi/4}\sqrt{\frac{k}{m^{*}}}\lb - \frac{e^{i\pi/4}}{\sqrt{m^{*}k}}  \partial_{x}^{2} +  \frac{\sqrt{m^{*}k}}{e^{i\pi / 4}} x^{2}\rb  - i  t/\ell^{2}  = \omega\lb - \partial_{z}^{2} + z^{2} \rb - it/\ell^{2}
\end{equation}
with $z \equiv e^{-i\pi/8 }\l k m^{*}\r^{1/4 } x$ and $\omega = e^{-i\pi/4}\sqrt{k/m^{*}}$. This operator can be factorized, as is typical, using the operators $\mathfrak{r} =  z- \partial_{z}$ and $\mathfrak{l} = z + \partial_{z}$. We note that $\mathfrak{r}\neq \mathfrak{l}^{\dagger}$; however, it is still true that $\lb \mathfrak{l}, \mathfrak{r}\rb =2$; $\l z^{2} - \partial_{z}^{2}\r e^{-z^{2}/2} = e^{-z^{2}/2}$; and $\mathfrak{l}e^{-z^{2}/2}=0$. Then, we can make use of the identity $z^{2} - \partial_{z}^{2} = \frac{1}{2} \lcb \mathfrak{r}, \mathfrak{l} \rcb$ to rewrite as. 
\begin{equation}
    \epsilon\l \hat{p}, \hat{x}\r = \frac{\omega}{2} \lcb \mathfrak{r}, \mathfrak{l}\rcb - it/\ell^{2} =  \omega\, \l  \mathfrak{r} \mathfrak{l}  + 1\r - it/\ell^{2}.
\end{equation}
This implies that we can produce a family of eigenstates $\psi_{n}\l z \r \propto  \l -1\r^{n}\mathfrak{r}^{n} e^{-z^{2}/2} = e^{-z^{2}/2} {\rm He}_{n}\l z \r$ with quasispectrum $\omega\l 2n + 1 \r - it/\ell^{2}$ (up to normalization; the minus sign is merely a convention), per the standard Harmonic oscillator ladder algebra. Our eigenstates are not, however, orthogonal to one another so it becomes extremely challenging to expand the initial conditions in this basis. 

We can instead recognize the key expression in the RHS of Eq.~\eqref{eqn:bound_start} as the sum of singular values squared, $\sum_{n} \sigma^{2}_{n}\big{(}\hat{U}^{t}\big{)}$. Applying Weyl's majorization theorem (see Theorem II.3.6 of \cite{bhatia2013matrix}), we can then lower bound the sum of the singular values squared with the sum of the absolute values of the eigenvalues squared:
\begin{eqnarray}
    && \sum_{x,y}\lv \bra{x} \hat{U}^{t} \ket{y}\rv^{2} = {\rm Tr}\lb \big( \hat{U}^{t}\big)^{\dagger}\hat{U}^{t}\rb =  \sum_{n=0}^{\infty}\sigma^{2}_{n}\big( \hat{U}^{t}\big)   \geq  \sum_{n=0}^{\infty} \lv \lambda^{2}_{n} \big( \hat{U}^{t}\big)\rv  \nonumber \\
    && = e^{-\sqrt{2k/m^{*}} t -2t^{2}/\ell^{2}}\sum_{n=0 }^{\infty} e^{-n\sqrt{8k/m^{*}} t } = \frac{e^{-\sqrt{2k/m^{*}} t - 2t^{2}/\ell^{2}} }{1 - e^{-\sqrt{8k/m^{*}} t }} > e^{- \sqrt{2k/m^{*}}t - 2t^{2}/\ell^{2}}
\end{eqnarray}
where we zero-index the singular values and eigenvalues for the ground state. Recalling that $k\sim t/\ell^{4}$ and inserting the above inequality into the RHS of Eq.~\eqref{eqn:bound_start}, we find norm-squared correlation function is lower bounded as
\begin{equation}
    \ell^{-1}\sum_{x,y} \lv \langle\sigma^{-}_{y}\l t \r \sigma^{+}_x\rangle\rv^{2}\gtrsim \ell^{-1} \exp\l- \mc{O}\l \ell + t^{2}/\ell^{2}+ t^{3/2}/\ell^{2}\r \r, 
\end{equation}
where we remind the reader that $p_{\ell} \sim e^{- \mc{O}\l \ell \r} $. Since the eigenstates have length scale $k^{-1/4} \sim \ell/t^{1/4} $ we expect the magnon to remain within the void to a parametrically good approximation. Thus, we find the effects of the subleading terms in the density profile are the confinement of the magnon to the void region and, recalling $\ell\sim t^{\alpha}$, the further suppression of the correlation function by a factor $\exp(-{\cal O}(t^{3/2-2\alpha}))$. The in-exponent contribution $t^{3/2-2\alpha}$ is parametrically smaller than the leading order contribution $t^{2-2\alpha}$, and therefore does not affect the asymptotic decay of the correlation function.

We expect higher order terms will also not matter by simple dimensional analysis: for the $x^{2m}$ term the coupling constant will scale $ t/\ell^{2\l m +1\r}$ and the length scale as $\ell/t^{1/4}$ so that the decay will be of order $t^{1 - m/2 - 2\alpha}$. Integrating with respect to time we find an overall contribution $t^{2-m/2-2\alpha}$ which, for $\alpha = 2/3$, tells us that the contribution from terms of higher order than $x^{2}$ vanish (even as subleading terms).

\section{Fluctuating nonlinear fluid with a single particle spectator}
In the previous section, we estimated the correlation function $\langle \sigma^-_{x}(t)\sigma^+_{0}(0)\rangle$ in terms of a magnon path integral which spectates a fluid density $n(x,t)$. In that section we restricted our attention to states which initially contain a void, and then annealed the fluid density. In the fully fluctuating case we upgrade the annealed density $n(x,t)$ to a fluctuating field $\hat{n}(x,t)$. This allows us to consider the contributions from rare density fluctuations, and also generalizes the previous section to generic initial density configurations. Although rare, such fluctuations could, in principle, provide a mechanism for a slower decay from a trade-off between probability of rare density patterns (other than those void configurations we have already considered) and the cost per collision with the inserted magnon. To address this question, we now consider the non-Hermitian evolution with a fully fluctuating density profile $\hat{n}(x,t)$ numerically.

Rather than solve the continuous time evolution in Eq.~\eqref{nhh}, we consider a stroboscopic evolution of this ``spectator" magnon, which evolves according to a single (quantum) particle propagator between timesteps, and which is spectator to a classical stochastic fluid with a filling dependent diffusivity $D(n)\sim 1/n$. The magnon wave-function $\psi(x,t)$ is dissipated stroboscopically at every timestep $\Delta t=1$. The dissipation now depends on the local occupation $\hat{n}(x,t)$ of a stochastic fluid, and is updated according to the rule $\psi(x,t) \to \exp(-\gamma \hat{n}(x,t))\psi(x,t)$. In practice, we store the magnon wavefunction as a vector of length $L$ (system size) and use a Bessel function propagator for the propagation of the magnon on the lattice $\psi_x(t) \to \sum_{x'} i^{{x-x'}}J_{x,x'}(\Delta t) \psi_{x'}(t)$ where $J_n(\tau)$ is the Bessel function of the first kind. This is the propagator associated with a single particle dispersion $\omega(k)=\cos(k)$, however, any (ballistic) single particle propagator would suffice. Simultaneous to the single-magnon propagation, we evolve the classical stochastic fluid for a time $\Delta t$, before finally applying the dissipation update rule. This sequence is then repeated for the full time evolution.

\begin{figure}
    \centering
    \includegraphics[width=0.45\linewidth]{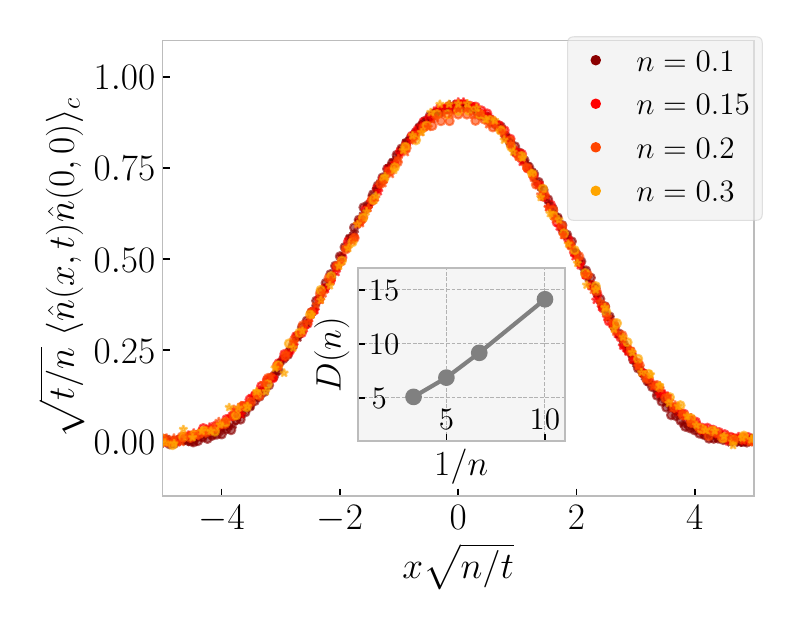}\ \raisebox{-0.015\height}{\includegraphics[width=0.45\linewidth]{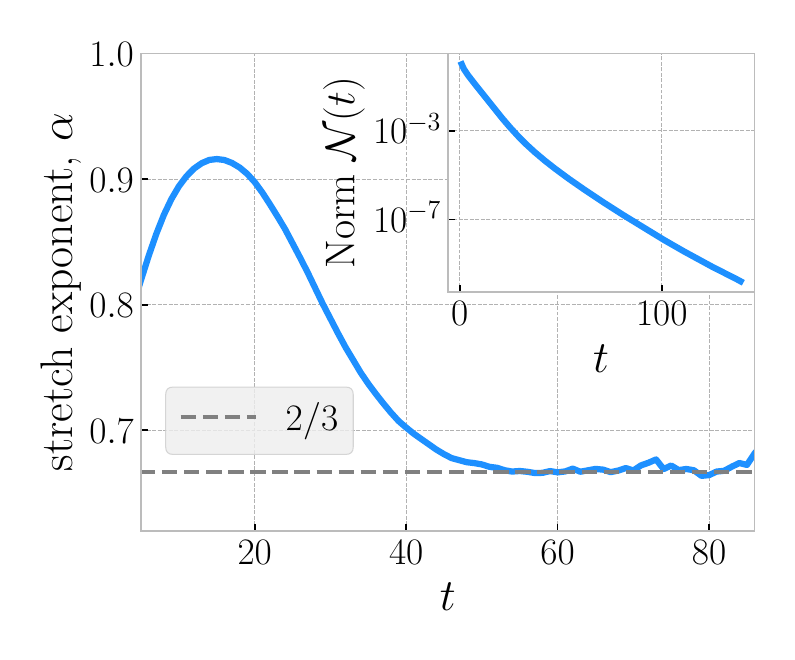}}
    \caption{(Left) The dynamical structure factor $\langle \hat{n}(x,t) \hat{n}(0,0)\rangle_{\text{conn}}$ for the classical stochastic gas described in text at several different particle densities $n$. We have rescaled space by $\sqrt{n/t}$ for a scaling collapse of the structure factor. (Inset) The diffusivity $D(n)$, computed using the Kubo formula \cite{kubo1957statistical}, shows linear dependence on $1/n$, the inverse of the particle density. (Right) The time-dependent stretch exponent $\alpha$ determined by a log derivative (Eq.~\eqref{eqn:log-derivative}) of the survival probability of a magnon undergoing dissipative dynamics while spectating the classical fluctuating fluid, as described in the text, showing good agreement with our analytic prediction $\alpha=2/3$. This data was obtained using Monte-Carlo sampling with $N=2\times 10^6$ samples. (Inset) The survival probability $P(t)$ is simply given by the norm ${\cal N}(t)$ of the single magnon wavefunction.}
    \label{fig:classical-gas-density-density-corr}
\end{figure}

We take this classical fluid to be a stochastic gas of a ballistic point particles with velocities $v = \sin(k)$ with uniform $k\in [-\pi,\pi)$, which are advected ballistically between collisions, and which randomize their velocities at each collision. Evaluating the fluid density-density correlation function $\langle \hat{n}(x,t) \hat{n}(0,0)\rangle_{\text{conn}} \equiv \langle \hat{n}(x,t) \hat{n}(0,0)\rangle - \langle \hat{n}(x,t) \rangle \langle \hat{n}(0,0)\rangle$, we confirm that this stochastic gas has a diffusivity $D(n)\sim 1/n$ by evaluating $D$ using the Kubo formula~\cite{kubo1957statistical}. The diffusive scaling collapse, and the diffusivity as a function of magnon density, is shown in Fig.~\ref{fig:classical-gas-density-density-corr} (left panel).
The survival probability $P(t)\equiv \langle \int dx |\psi(x,t)|^2\rangle_{\hat{n}}$ of the spectator magnon (averaged over the stochastic evolution of the classical gas) is computed numerically, using Monte-Carlo sampling with $N=2\times 10^6$ samples. This is shown in Fig.~\ref{fig:classical-gas-density-density-corr} (right panel), along with the time-dependent stretch exponent $\alpha(t)$ defined using a log derivative,
\begin{equation}\label{eqn:log-derivative}
   \alpha(t) \equiv \frac{d\log{-\log{ P(t)}}}{d\log{t}}.
\end{equation}
The numerically obtained stretch exponent is seen to plateau at $\alpha=2/3$ at long times, in good agreement with the claimed stretched exponential decay in the main text.

\section{Additional Floquet numerics}

\begin{figure}
    \centering
    \includegraphics[width=0.3\linewidth]{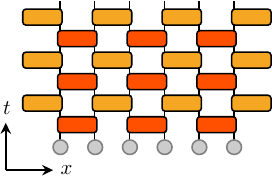}
    \caption{Translationally invariant Floquet circuit with a Floquet period comprised of a single application of an even and odd layer of  two-site gates. Unitary gates of the same color are identical.}
    \label{fig:Floquet-circuit}
\end{figure}

In this section we present numerical data for three different translationally invariant Floquet models, the first of which is also presented in Fig.~\ref{fig:2} in the main text. We consider translationally invariant Floquet circuits where a single Floquet period is composed of an even and odd layer of two-site gates, $U_F = U_e U_o$, where $U_e = \prod_{n=1}^{L/2}U_{2n-2,2n-1}$ and $U_o = \prod_{n=1}^{L/2}U_{2n-1,2n}$. This circuit has translational invariance under translations of $2m$ sites $m\in \mathbb{Z}$, as can be seen in Fig.~\ref{fig:Floquet-circuit}. We choose the unitary gates to be given by $U_{x,x+1} = \exp(iH_{x,x+1})$ where $H_{x,x+1}$ is an anisotropic Heisenberg interaction with a staggered field (to break integrability~\cite{MarkoIntegrableCircuits}) and staggered anisotropy,
\begin{equation}\label{eqn:model_A}
    H_{x,x+1} = J(\sigma^x_{x}\sigma^x_{x+1} + \sigma^y_{x}\sigma^y_{x+1})+\Delta\Big(1+\delta(-1)^x\Big) \sigma^z_x \sigma^z_{x+1} +(-1)^j g (\sigma^z_x - \sigma^z_{x+1}),
\end{equation}
where $\sigma^{\alpha=x,y,z}$ are Pauli matrices. For the first model (A) we take $(J,\Delta,\delta,g) = (0.393,0.177,0.333,0.3)$; for the second model (B) we take $(J,\Delta,\delta,g) = (0.393,0.293,0,0.2)$; for the third model (C) we take $(J,\Delta,\delta,g) = (0.589,0.514,0,0.45)$.
These models are chosen to be non-integrable, to have a single conserved quantity, and to be diffusive at $O(1)$ filling (see next section), and are otherwise not fine-tuned. When models exhibit reliable diffusion at ${\cal O}(1)$ filling, we find a stretched exponential decay of non-hydrodynamic correlation functions, as shown in the final section.

\subsection{Diffusive hydrodynamic correlations at ${\cal O}(1)$ filling}

\begin{figure}
    \centering
    \includegraphics[width=0.85\linewidth]{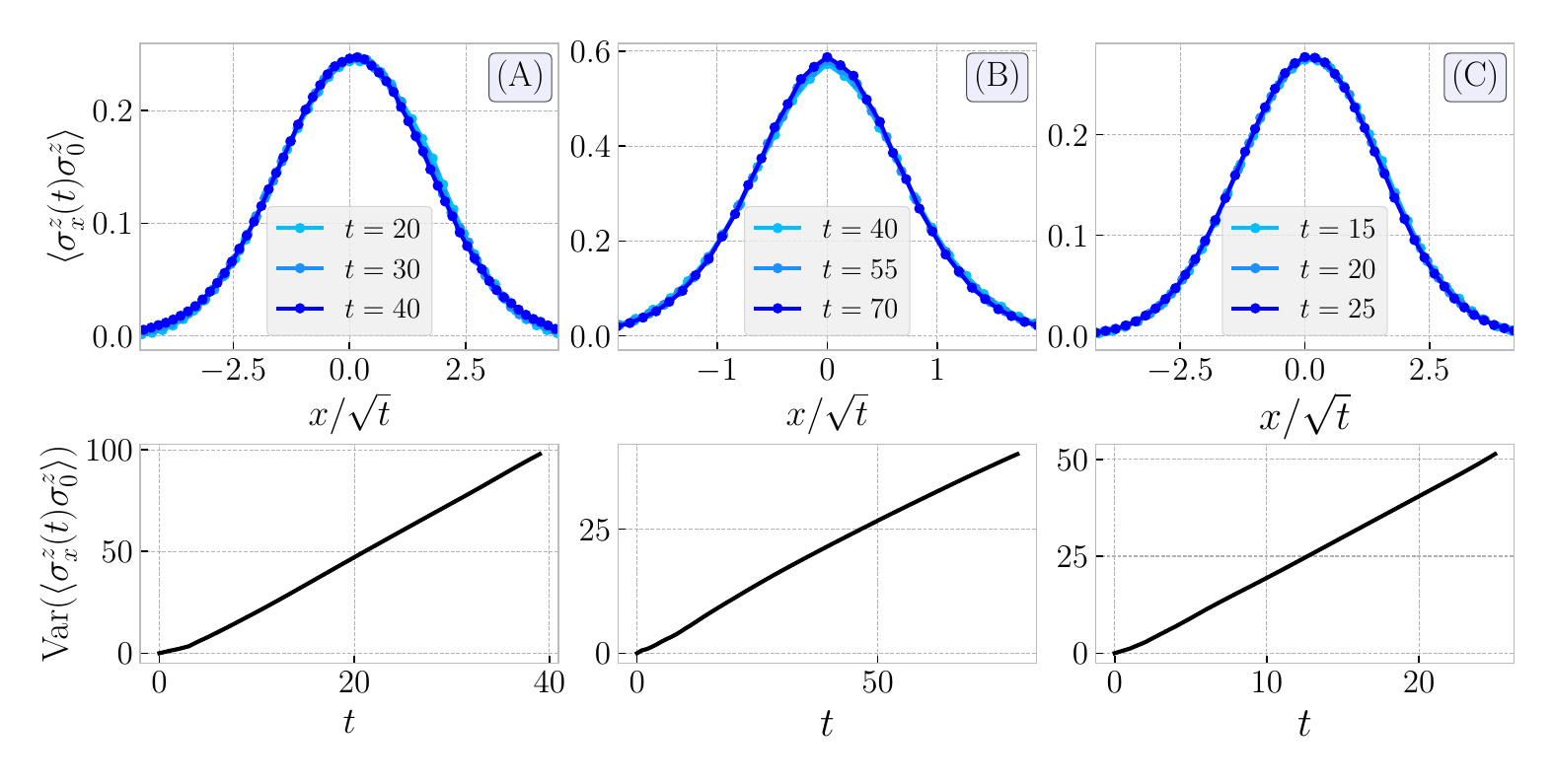}
    \caption{Hydrodynamic correlations in three chaotic U$(1)$-conserving translationally invariant Floquet circuits. (Top) The infinite temperature dynamical structure factor $\langle \sigma^z_x(t) \sigma^z_0 \rangle_c$, shows a diffusive scaling collapse in each case. (Bottom) The mean-squared displacement $\text{Var}(\langle \sigma^z_x(t) \sigma^z_0 \rangle_c) = \sum_{x} x^2 \langle \sigma^z_x(t) \sigma^z_0 \rangle_c$ showing linear growth, consistent with diffusion. From left to right we show model A, B, and C, as described in the text.}
    \label{fig:Floquet_ZZ_corr}
\end{figure}

Each of these models has only a single conserved quantity, total magnetization $\sum_j \sigma^z_j$, which we verify is diffusive at half filling (infinite temperature) by TEBD simulations measuring the dynamical structure factor $\langle \sigma^z_x(t) \sigma^z_0 \rangle_c \equiv \langle \sigma^z_x(t) \sigma^z_0 \rangle - \langle \sigma^z_x(t)\rangle \langle \sigma^z_0 \rangle$ on systems of size $L=60$ and with a maximum bond-dimension $\chi=500$. This data is shown in Fig.~\ref{fig:Floquet_ZZ_corr}, along with the mean-squared displacement, $\text{Var}(\langle \sigma^z_x(t) \sigma^z_0 \rangle_c) = \sum_{x} x^2 \langle \sigma^z_x(t) \sigma^z_0 \rangle_c$, which is seen to grow linearly in time.

\subsection{Non-hydrodynamic correlation functions at low densities}
\begin{figure}
    \centering
    \includegraphics[width=0.9\linewidth]{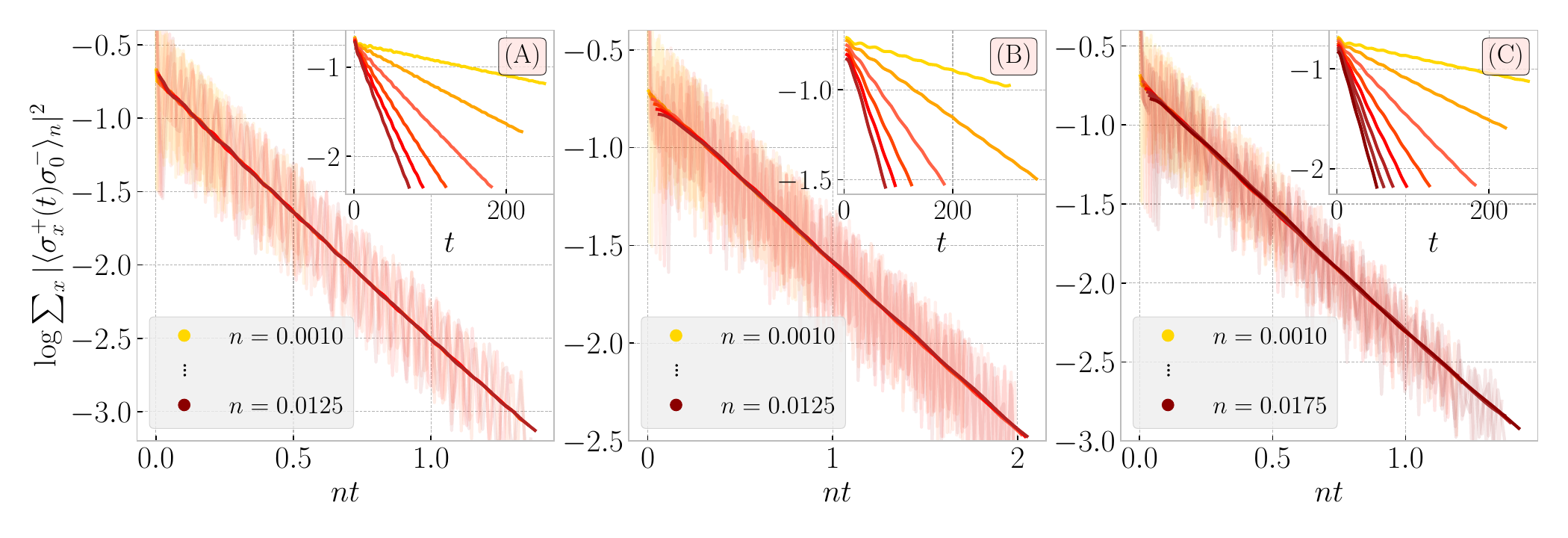}
    \caption{The non-hydrodynamic correlation function $\sum_x|\langle \sigma_x^+(t) \sigma^-_0\rangle_n|^2$ evaluated at low densities $n \in [0.001, 0.0175]$. (Main panels) The logarithm of the correlation function plotted against $n t$ showing the expected scaling collapse. The dark curves show data smoothed by convolving with a Gaussian kernel over a timescale $\Delta t=2.5$ (in units of the Floquet period); the light curves show the raw data. (Insets) The Gaussian smoothed data shown against time (without rescaling). From left to right we show model A, B, and C, as described in the text.}
    \label{fig:Floquet-low-den-supp-mat}
\end{figure}

A main ingredient of the stretched exponential decay of non-hydrodynamic correlation functions in translationally invariant Floquet systems was the vanishing decay rate in the low density limit $n \to 0$. In particular, we observed that for times before the mean-free-time, $t\sim 1/n$, the decay rate is equal to the magnon density $n$, which we attributed to the rate of dephasing events due to collisions. While we showed only a single model (A) in the main text, we now demonstrate the generality of this result by considering two other chaotic translationally invariant Floquet models. In Fig.~\ref{fig:Floquet-low-den-supp-mat} we show the correlation function $\langle \sigma_x^+(t) \sigma^-_0\rangle_n$ (squared and summed over $x$) decays with a rate proportional to $\rho$ in all three models (A, B and C) introduced previously. The expectation value $\langle O \rangle_n\equiv \Tr(\rho_{\mu(n)} O )$ is with respect to a (infinite temperature) Gibbs state $\rho_{\mu(n)}$ with chemical potential $\mu(n)$ corresponding a magnon density $n$. This data is obtained by TEBD simulations with bond dimension $\chi = 200$.

We point out that the simulated times are only ${\cal O}(1)$ in units of the mean-free-time. In this regime, the dynamics is purely ballistic, and the void mechanism is not able to slow the decay of non-hydrodynamic correlations. Voids of length $\ell$ are melted on a timescale $\tau\sim \ell$, meaning that a void of length $t$ is required to shield the inserted magnon from a ballistic (${\cal O}(t)$) number of collisions by time $t$. The cost of such a void is $\exp(-{\cal O}(nt))$, the same as the conjectured cost due to dephasing from collisions in a typical configuration of magnons. Therefore, by probing the low density limit before the mean-free-time, we are able to work in a regime in which rare void configurations do not dominate over typical charge configurations, and the decay of the correlation function should be exponential with a rate proportional to the density. The low density correlation functions for the three example models confirm this (see Fig.~\ref{fig:Floquet-low-den-supp-mat}) and are clear evidence of dephasing by collisions as a mechanism for correlation decay.

\subsection{Non-hydrodynamic correlations at infinite temperature}
\begin{figure}
    \centering
    \includegraphics[width=0.95\linewidth]{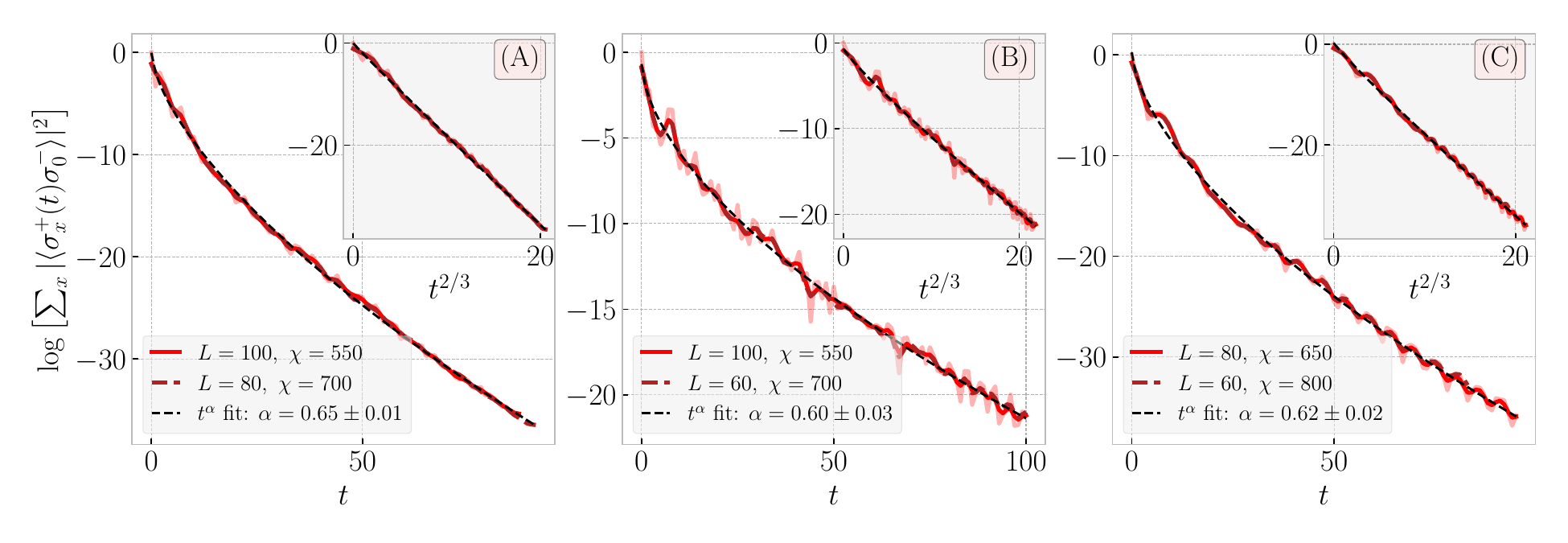}
    \caption{The non-hydrodynamic correlation function $\sum_x|\langle \sigma_x^+(t) \sigma^-_0\rangle|^2$ evaluated at infinite temperature. (Main panels) The logarithm of the correlation function. The dark curves show data smoothed by convolving with a Gaussian kernel over a timescale $\Delta t=1$ (in units of the Floquet period); the light curves show the noisier raw data. The dashed black curve is an power law fit ${\cal O}(t^\alpha)$ to the (logarithm of the) correlation data. (Insets) The Gaussian smoothed data shown against a squashed time-axis, $t^{2/3}$. From left to right we show model A, B, and C, as described in the text.}
    \label{fig:Floquet-infT-spsm-supp-mat}
\end{figure}

In this section we present evidence that the stretched exponential decay of non-hydrodynamic correlation functions is generic by considering the correlation function $\langle \sigma^+_x(t) \sigma^-_0\rangle$ in three chaotic translationally invariant U$(1)$-conserving Floquet models (models A, B and C) introduced previously. We have just seen that these models are diffusive at half-filling and have the expected correlation decay at low densities. Therefore, the correlation function at ${\cal O}(1)$ filling should decay as a stretched exponential in each of these models, with a stretch exponent $\alpha\leq 2/3$, as argued in the main text. Using TEBD simulations, we directly simulate the operator evolution of $\sigma^+_0$ for each of these models up to times $t\approx 100$ (in units of the Floquet period) and find good agreement between this bound and the fitted stretch exponent: $\alpha=0.65\pm 0.01$ for model A, $\alpha=0.60\pm 0.03$ in model B, and $\alpha=0.62\pm 0.02$ in model C. The correlation function data and fits are shown in Fig.~\ref{fig:Floquet-infT-spsm-supp-mat}.

\end{document}